\documentclass[aps,prb,superscriptaddress,showkeys,reprint]{revtex4-2}

\usepackage{amsmath}
\usepackage{mathtools}
\usepackage{subdepth}
\usepackage{physics}
\usepackage{xfrac}
\usepackage{txfonts}
\usepackage{stmaryrd}
\usepackage[T1]{fontenc}
\usepackage{graphicx}
\usepackage{bm}
\usepackage{orcidlink}
\usepackage{hyperref}
\definecolor{linkblue}{HTML}{2E3092}
\hypersetup{
colorlinks = true,
linkcolor  = linkblue,
citecolor  = linkblue,
urlcolor   = linkblue,
filecolor  = linkblue
}
\usepackage{mleftright}
\mleftright
\renewcommand{\Re}{\mathfrak{Re}}
\renewcommand{\Im}{\mathfrak{Im}}
\begin{document}
\title{Invariant-based master equation applied to driven qutrit coupled to a bath and a leaky cavity}
\author{Sagarika Basak\,\orcidlink{0000-0003-2069-644X}}
\email{basak.sagarika@ou.edu}
\affiliation{Homer L. Dodge Department of Physics and Astronomy,\\\href{https://ror.org/02aqsxs83}{The University of Oklahoma}, 440 W. Brooks Street, Norman, Oklahoma 73019, USA}
\affiliation{Center for Quantum Research and Technology,\\\href{https://ror.org/02aqsxs83}{The University of Oklahoma}, 440 W. Brooks Street, Norman, Oklahoma 73019, USA}

\author{A.~Javadi\,\orcidlink{0000-0002-8833-0738}}
\affiliation{Homer L. Dodge Department of Physics and Astronomy,\\\href{https://ror.org/02aqsxs83}{The University of Oklahoma}, 440 W. Brooks Street, Norman, Oklahoma 73019, USA}
\affiliation{Center for Quantum Research and Technology,\\\href{https://ror.org/02aqsxs83}{The University of Oklahoma}, 440 W. Brooks Street, Norman, Oklahoma 73019, USA}
\affiliation{School of Electrical and Computer Engineering,\\\href{https://ror.org/02aqsxs83}{The University of Oklahoma}, 110 W. Boyd Street, Norman, Oklahoma 73019, USA}

\author{D.~Blume\,\orcidlink{0000-0001-7381-5698}}
\email{doerte.blume-1@ou.edu}
\affiliation{Homer L. Dodge Department of Physics and Astronomy,\\\href{https://ror.org/02aqsxs83}{The University of Oklahoma}, 440 W. Brooks Street, Norman, Oklahoma 73019, USA}
\affiliation{Center for Quantum Research and Technology,\\\href{https://ror.org/02aqsxs83}{The University of Oklahoma}, 440 W. Brooks Street, Norman, Oklahoma 73019, USA}
\date{\today}

\begin{abstract}
We employ a generalized approach to the master equation for driven open $N$-level ($N>2$) quantum systems using Lewis-Riesenfeld invariants, which avoids the driving-strength restrictions inherent to conventional approaches. 
We show that the invariant-based master equation provides a unifying generalized framework, which reduces to the frequently employed laboratory-frame master equations and the less frequently employed rotating-frame master equation framework under appropriate simplifications.
Extending the prototypical two-level system, we show that the inclusion of another state coupled to the ground state via reservoir-induced dephasing gives rise to qualitatively new dissipative behaviors that are, in general, not captured by standard approximations. We also apply the invariant-based master equation framework to a driven quantum dot coupled to a leaky cavity, demonstrating the framework's ability to capture relevant dissipative dynamics without additional assumptions. Our work paves the way for quantum-control applications in the presence of dissipation.
\end{abstract}

\maketitle
\section{\label{sec:introduction}Introduction}
Reliable theoretical modeling of externally driven quantum dynamics is essential for the development of quantum devices, quantum simulators, and quantum computers. When the system is not isolated, the degrees of freedom of the environment are frequently accounted for effectively within a master equation formulation~\cite{breuer2007,Rivas2012}. While the reduction to the system degrees of freedom is in many cases the only feasible—and often quite successful---avenue for treating the dynamics, it is well-known that this strategy is plagued by fundamental and practical issues~\cite{breuer2007,fernandezdelapradilla2024}. Typical time-dependent master equations that go beyond the strict Lindblad form are, e.g., incompatible with fundamental principles of both quantum mechanics and thermodynamics such as complete positivity (negative populations may arise)~\cite{dann2018,hartmann2020,dabbruzzo2023,fernandezdelapradilla2024} and the second law of thermodynamics (entropy may decrease)~\cite{tupkary2022,tupkary2023,fernandezdelapradilla2024}.

Derivations of master equations can be, broadly, grouped into two categories~\cite{breuer2007}. The first may be best described as macroscopic or axiomatic. The second, pursued in this work, is microscopic. Numerous variants exist within this latter category, with key differences in the approximations made and correspondingly the complexity and applicability regime of the resulting master equation~\cite{breuer2007,jeske2015,eastham2016,hartmann2020,mccauley2020,davidovic2020,farina2019,nathan2020,fernandezdelapradilla2024,Coleman2022}. This work builds on a Lie-algebra- or dynamical invariant-based master equation (IME) framework that has, so far, only been applied to a few paradigmatic systems, such as the harmonic oscillator and the two-level system, and allows for higher-order terms to be accounted for systematically order-by-order~\cite{dann2018,boubakour2025,wu2022,BasakInPrep}. Dynamical or Lewis-Riesenfeld invariants are conserved quantities that are inherently tied to an underlying symmetry~\cite{lewis1969,xichen2011}. While this requirement might appear restrictive, it is important to note that the dynamical invariant $I_q(t)$, which is being utilized in our framework, is a characteristic of the isolated system Hamiltonian (in the absence of the environment), thereby making it broadly applicable. Leveraging the eigenstates of $I_q(t)$, the approach bypasses time-ordering issues, which lie at the heart of several key approximations made in typical master equation derivations~\cite{shavit2019,fernandezdelapradilla2024}.

\begin{figure*}[t]
\includegraphics[width=6.75in]{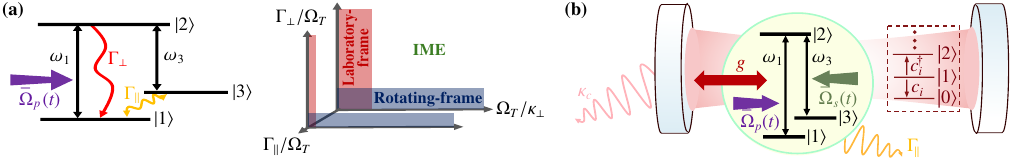}
\caption{\label{fig:System}(color online) We consider a qutrit with states $\ket{1},\,\ket{2},\,\text{and }\ket{3}$. The $(1{\leftrightarrow}2)$-transition (transition frequency $\omega_1$) is coupled by a field with drive strength $\Omega_{p}$ and drive frequency $\omega_{p}$. The $(2{\leftrightarrow}3)$-transition frequency is denoted by $\omega_3$. (a) Application~1: The qutrit is coupled to a thermal bath via transverse coupling that induces dissipation from $\ket{2}$ to $\ket{1}$ (decay rate $\Gamma_{\perp}$) and longitudinal coupling that induces dephasing between $\ket{3} \text{~and~} \ket{1}$ (decay rate $\Gamma_{\parallel}$). The latter introduces a new scale compared to the driven 2-level system. The validity regimes of conventional laboratory- and rotating-frame approaches, summarized in $\Gamma_{\perp}/\Omega_T$ versus $\Omega_T/\kappa_{\perp}$ space for the 2-level system~\cite{shavit2019}, are reduced if one moves away from zero along the $\Gamma_{\parallel}/\Omega_T$ axis; $\Omega_T$ denotes a generalized Rabi coupling strength. (b) Application~2: In contrast to Application~1, the $(2{\leftrightarrow}3)$-transition is also driven, with drive strength $\Omega_{s}$ and drive frequency $\omega_{s}$. The three-level quantum dot is coupled to a single-mode cavity with coupling strength $g$. The quantum dot and cavity are coupled to a thermal bath with qutrit dephasing rate $\Gamma_{\parallel}$ and cavity decay rate $\kappa_c$, respectively. The latter sets the effective transverse decay rate $\Gamma_{\perp}$ and spectral width of the transverse coupling $\kappa_{\perp}\,(\Gamma_{\perp} = 2g^2/\kappa_c$ and $\kappa_{\perp} = \kappa_c)$, which emerge when treating the cavity as a bath.}
\end{figure*}

To demonstrate the practical utility of the IME framework for $N$-level systems with $N>2$, we consider a qutrit coupled to a bath with Lorentzian spectral function. In Application~1, we consider the simplest nontrivial extension of a two-level system, namely, we consider the situation where the third level is coupled dissipatively (dephasing rate $\Gamma_{\parallel}$) to the ground state of the driven Rabi-coupled two-level system [levels 1 and 2; Fig.~\ref{fig:System}(a)]. It is demonstrated that a non-vanishing $\Gamma_{\parallel}$ reduces the validity regime of so-called laboratory-frame and rotating-frame master equations, which have been applied to the two-level system~\cite{shavit2019}. A comparative analysis of the fluorescence spectrum $S_{12}(\omega)$, determined via the invariant-based, laboratory-frame, and rotating-frame master equations, shows that the addition of the third dissipatively coupled state introduces qualitatively new features.
In Application~2, the third level is not only dissipatively coupled to the first level but also coherently to the second level via the time-dependent Rabi coupling $\overline{\Omega}_{s}(t)$ [Fig.~\ref{fig:System}(b)]. This set-up is directly applicable to a three-level quantum dot that is embedded into a leaky single-mode cavity, which is, in turn, coupled to the environment. In an effective description, the cavity and environment serve as baths with Lorentzian spectral functions. It is shown that the IME predicts, for parameter combinations that can be realized in state-of-the-art experiments, distinct features in the fluorescence spectra that are not captured by either the laboratory-frame or rotating-frame master equations, including modifications of the so-called Mollow triplets, which have been observed experimentally in quantum dots~\cite{ulrich2011,muller2007,konthasinghe2012,ulhaq2012,nickvamivakas2009,flagg2009,ates2009}, NV centers~\cite{wang2021}, and cold atoms~\cite{Ng2022,schuda1974,hartig1976,grove1977,ortiz-gutierrez2019}.

The derivation and applications of the IME to the three-level system presented in this work have broad implications beyond the two examples presented in this work. We show that the IME framework reduces, under appropriate simplifications, to more approximate descriptions such as the frequently employed laboratory-frame master equation framework and the less frequently employed rotating-frame master equation framework. Our work thus shows that the IME provides a much needed unifying framework that reduces to known limiting descriptions. Moreover, our work suggests that there exist no fundamental roadblocks to applying the IME to driven $N$-level systems, with $N>3$. The approach accommodates arbitrary time-dependent driving, for which Floquet theory~\cite{Mosallanejad2025,alicki2012,levy2012} may be inapplicable, and allows for systematic order-by-order improvements. Our framework opens the door for theoretically describing state preparation and quantum control protocols of driven systems with competing scales, as encountered in quantum hybrid systems.

The remainder of this paper is organized as follows. Section~\ref{sec:thesystem} introduces the system under study. Starting with the Redfield master equation, the IME framework is developed in Sec.~\ref{sec:masterequation}. Section~\ref{sec:definition} defines the fluorescence spectrum and provides a brief review of key features of the Mollow triplets for a two-level system. The fluorescence spectra of the driven qutrit are discussed in Sec.~\ref{sec:results}. Finally, Sec.~\ref{sec:summary} provides a summary and outlook. Details of the technical derivations and additional background information are relegated to Appendices~\ref{sec:appendA}--\ref{sec:appendE}.

\section{\label{sec:thesystem}System under study}
Throughout, we use units where $\hbar=1$. In the lab frame, the total Hamiltonian $H$ is written as a sum of the time-dependent system term $H_{q}(t)$, the bath term $H_b$, and the system--bath interaction term $H_{I}$, $H=H_{q}(t)+H_b+H_I$. The system Hamiltonian comprises the time-independent $H_q^0$ that describes a qutrit ($\Lambda$-system with ground states $\ket{1} \text{ and } \ket{3}$ and excited state $\ket{2}$) and the time-dependent drive $H_q^D(t)$, $H_q(t) = H_q^0 + H_q^D(t)$, where
\begin{equation}
\begin{split}
\label{eq_quantumdot}
    H_q^0 &= -\omega_1 \ket{1}\bra{1} -\omega_3 \ket{3}\bra{3} \,, 
    \\H_q^D(t) &= \overline{\Omega}_p(t) \left( \ket{2}\bra{1} + \ket{1}\bra{2}\right) +  \overline{\Omega}_{s\vphantom{p}}(t)\left( \ket{2}\bra{3} + \ket{3}\bra{2}\right)\,.
\end{split}
\end{equation}
Here, $\omega_1 \text{~and~} \omega_3$ are the transition frequencies between $\ket{1} \text{~and~} \ket{2}$ and between $\ket{3} \text{~and~} \ket{2}$, respectively, and $ \overline{\Omega}_p(t) \text{~and~} \overline{\Omega}_s(t)$ are the strengths of the fields that couple these states  (Fig.~\ref{fig:System}). We use 
\begin{eqnarray}
 \overline{\Omega}_p(t)=\Omega_{p} \cos(\omega_p t)
\end{eqnarray}
and 
\begin{eqnarray}
 \overline{\Omega}_s(t)=\Omega_{s} \cos(\omega_s t)\,,
\end{eqnarray}
where the Rabi frequencies $\Omega_p \text{~and~} \Omega_s$ are real and $\omega_p \text{~and~} \omega_s$ denote driving frequencies. We emphasize that the formalism can also be applied to non-periodic drives. For simplicity, the examples considered in this work utilize periodic drives; applications to non-periodic drives will be considered in follow-up work.

For the analysis below, it is convenient to work in the rotating frame. The transformation from the lab frame to the rotating frame and the corresponding rotating-frame Hamiltonian are given in Appendix~\ref{sec:appendA}. Notice that Application 1 does not consider any additional coherent drives beyond that which couples, as in a two-level system, states $\ket{1} \text{~and~} \ket{2}$.

The bath Hamiltonian is characterized by mode frequencies $\omega_{b,k}$, 
\begin{eqnarray}
\label{eq_ham_bath}
    H_b=\sum_k \omega_{b,k} b_k^{\dagger} b_k\,,
\end{eqnarray}
with the $b_k^{\dagger}$ denoting bosonic operators that excite the $k$th bath mode. The system--bath interaction accounts for both transverse and longitudinal decoherence, with coupling constants $g_{\perp,k} \text{~and~} g_{\parallel,k}$, respectively, 
\begin{equation}
H_I
= \displaystyle\sum\limits_k
\left[
g_{\perp,k}\,\sigma_{12}
+\dfrac{g_{\parallel,k}}{2}
\left(\sigma_{33}-\sigma_{11}\right)
+ \text{H.c.}
\right] \left(b_k^{\dagger}+b_k\right)
\,,
\label{eq_intham}
\end{equation}
where $\sigma_{mn}=\ket{m}\bra{n}$.
Throughout, we assume that the bath consists of a continuum of modes that is characterized by Lorentzian spectral density functions $J_{\beta}(\omega)$ $\left[\sum_k \abs{g_{\beta,k}}^2\rightarrow \int d\omega J_{\beta}(\omega)\right]$ with widths $\kappa_{\beta}$ centered at $\omega_{\beta}$,   
\begin{equation}
    J_{\beta}(\omega) = \dfrac{\Gamma_{\beta}}{\pi}\left[\dfrac{\left(\kappa_{\beta}/2\right)^2}{\left(\kappa_{\beta}/2\right)^2 + \left(\omega-\omega_{\beta}\right)^2}\right]\,.
\end{equation}
Here, $\beta$ stands for $\perp$ or $\parallel$, i.e., we assume separate spectral functions for the dissipative transverse and longitudinal processes. The derivation of the IME considered below uses the Redfield master equation \cite{breuer2007} and makes the ``standard'' Born-Markov approximation, which requires $g_{\beta,k}\ll \omega_1,\omega_3$ and $1/\kappa_{\beta} \ll 1/\max_k(g_{\beta,k})$, i.e., the bath is assumed to be weakly coupled to the system and assumed to reach its equilibrium on time scales much smaller than those for the system--bath couplings. A key point of the IME is that it probes the spectral functions $J_{\beta}(\omega)$ at the correct frequencies, i.e., at drive-dependent $\omega$ values. 

\section{\label{sec:masterequation}General invariant-based master equation framework}
Denoting the density matrices of the driven three-level system in the Schr\"odinger and interaction pictures by 
$\rho_q(t) \text{~and~} \tilde{\rho}_q(t)$, respectively, and the time-independent density matrix of the 
bath (same in the two pictures) by $\rho_b$, the Redfield master equation reads \cite{breuer2007} (see also Appendix~\ref{sec:appendB})
\begin{align}
\label{eq_master1}
\begin{split}
    \frac{d}{dt}\tilde{\rho}_q(t) =& - \int\limits_{0}^{\infty} ds\text{Tr}_b \left\{ \comm{\tilde{H}_{I}(t)}{\comm{\tilde{H}_{I}(t-s)}{\tilde{\rho}_q(t)  \otimes  \rho_b}} \right\}\,.
\end{split}
\end{align}
The Redfield master equation is, in general, non-Lindbladian. This implies, as discussed further below, that it allows, in principle, for (unphysical) negative populations~\cite{fernandezdelapradilla2024}.
Equation~(\ref{eq_master1}) can be simplified by expressing the interaction Hamiltonian as a tensor product of operators $\tilde{A}_{\beta}(t) \text{~and~} \tilde{B}(t)$ that live in the Hilbert spaces of the three-level system and the bath, respectively, 
\begin{eqnarray}
    \tilde{H}_{I}(t) = \textstyle\sum\limits_{\beta=\perp,\parallel}\tilde{A}_{\beta}(t) \otimes \tilde{B}(t)\,.
\end{eqnarray}
Here, the tilde denotes operators in the interaction picture.
For example, we have 
\begin{eqnarray}
    \tilde{A}_{\beta}(t)=U_q^{\dagger}(t)A_{\beta}U_q(t)
\end{eqnarray}
and 
\begin{eqnarray}
    \tilde{B}(t)=U_b^{\dagger}(t)BU_b(t)\,,
\end{eqnarray}
where 
\begin{eqnarray}
U_q(t) = \mathcal{T}e^{-i\textstyle\int_{0}^{t} H_q(s) ds}
\end{eqnarray}
and
\begin{eqnarray}
    U_b(t) = e^{-iH_b t}\,; 
\end{eqnarray}
$\mathcal{T}$ denotes the time ordering operator.
Defining 
\begin{eqnarray}
    \Lambda(s) = \text{Tr}_b \{\tilde{B}(t)\tilde{B}(t-s)\rho_b\}\,,
\end{eqnarray}
Eq.~(\ref{eq_master1}) becomes (see Appendix~\ref{sec:appendB})
\begin{widetext}
\vspace{-1\baselineskip}
\begin{align}
\label{eq_master2}
     \begin{split}
     \dfrac{d }{dt}\tilde{\rho}_q(t) = \textstyle\sum\limits_{\beta = \perp,\parallel}
     {\textstyle\int\limits_{0}^{\infty}} ds \left\{
     \left[\tilde{A}_{\beta}(t-s)\tilde{\rho}_q(t)\tilde{A}_{\beta}(t)-\tilde{A}_{\beta}(t)\tilde{A}_{\beta}(t-s)\tilde{\rho}_q(t) \right]\Lambda(s)
     +
     \left[ \tilde{A}_{\beta}(t)\tilde{\rho}_q(t) \tilde{A}_{\beta}(t-s)-\tilde{\rho}_q(t)\tilde{A}_{\beta}(t-s) \tilde{A}_{\beta}(t) \right]\Lambda^*(s)\right\}\,. 
     \end{split}
\end{align}
\vspace{-1\baselineskip}
\end{widetext}

To proceed, the integrand of Eq.~(\ref{eq_master2}) needs to be recast in a tractable form. This is, in general, a challenging task, as the integrand involves operators that are being evaluated at different times. As such, the next step in the derivation typically involves a series of approximations, ranging from restricting the driving amplitude to neglecting certain time-dependent terms. Proceeding without making any approximations, we express the rotation operator $U_q(t)$ in terms of the eigenstates $|\mu_n(t)\rangle$ of the invariant $I_q(t)$, which form an orthonormal basis, and the invariant phases $\alpha_n(t)$, 
\begin{equation}
\label{eq_propagator}
    U_q(t) = \textstyle\sum_n e^{i\alpha_n(t)}|\mu_n(t)\rangle\langle\mu_n(0)|\,.
\end{equation}
The invariant is defined through 
\begin{equation}
    \label{eq_def_iq}
    \dfrac{\partial}{\partial t} I_q(t) = i\comm{I_q(t)}{H_q(t)}\,
\end{equation}
and the invariant phases through
\begin{equation}
    \alpha_n(t) = \textstyle\int_{0}^{t}\expval{i \dfrac{\partial}{\partial \tau}-H_q(\tau)}{\mu_n(\tau)}d\tau\,.
\end{equation}
From the definitions it is evident that the invariant is strictly obtained from the system Hamiltonian $H_q(t)$ and does not require any information of the reservoir or the decoherence processes. Note that Eq.~(\ref{eq_def_iq}) defines not a unique $I_q(t)$ but a family of Hermitian operators. This flexibility can be used to ``absorb'' a varying amount of time dependence in $U_q(t)$. 

For the system under consideration, in the rotating frame and under the rotating wave approximation, the eigenstates of the invariant become time-independent and are given by (see Appendix~\ref{sec:appendA}) 
\begin{align}
\label{eq_iq_eb}
|\mu_0\rangle
&= \cos\zeta\ket{1} - \sin\zeta\ket{3}\,, \nonumber
\\
|\mu_+\rangle
&= \sin\phi \sin\zeta \ket{1} 
 + \sin \phi\cos\zeta \ket{3} + \cos\phi \ket{2}\,, \nonumber
\\|\mu_-\rangle
&= \cos\phi \sin\zeta \ket{1} 
 + \cos \phi\cos\zeta \ket{3} - \sin\phi \ket{2}\,,
\end{align}
where
\begin{align}
\zeta = \tan^{-1}\left(\Omega_p / \Omega_s\right), ~
\phi  = \frac{1}{2}\tan^{-1}\left(\sqrt{\Omega^2_s+\Omega^2_p}/\Delta\right)
\,.
\end{align}
The quantity $\Delta$ denotes the detuning of the field from the transition frequencies, $\Delta = \omega_1-\omega_p = \omega_3-\omega_s$. 
Using Eq.~(\ref{eq_iq_eb}) and the system Hamiltonian in the rotating frame, the phases reduce to 
\begin{align}
\alpha_0(t) 
= 
t \Delta , ~
\alpha_{\pm}(t) 
= \left(\dfrac{\Delta}{2}\mp \dfrac{1}{2}\sqrt{\Delta^2 + \Omega^2_s+\Omega^2_p}\right)t\,.
\end{align}
Using the invariant eigenbasis $\{|\mu_n\rangle\}$ to rewrite the system operators $\tilde{A}_{\beta}(t)$ in the interaction picture, the jump operators are independent of time in the interaction and Schr\"odinger pictures. This feature is critical and allows us to proceed without additional assumptions. Simplifying and transforming to the Schr\"odinger picture in the rotating frame (see {Appendix}~\ref{sec:appendB}), we find
\begin{eqnarray}
\label{eq_master3}
     \dfrac{d}{dt}\rho_{q,R}(t) 
     = 
     -i\comm{H_{q,R}}{\rho_{q,R}(t)} 
       +  \nonumber \\
       \sum_{mn,m'n'} \Gamma_{mn,m'n'}(t) \left[     
        F_{m'n'}\rho_{q,R}(t)F^{\dagger}_{mn}
       -F^{\dagger}_{mn}F_{m'n'}\rho_{q,R}(t)  
     \right]+ \text{H.c.}\,, \nonumber \\
\end{eqnarray}
where the jump operators $F_{mn}$ are defined as 
\begin{equation}
\label{eq_jop}
    F_{mn} = |\mu_m\rangle\langle\mu_n|\,.
\end{equation}
Explicit expressions for the effective time-dependent dissipation coefficients $\Gamma_{mn,m'n'}(t)$ can be found in Appendix~\ref{sec:appendB}.
While Eqs.~(\ref{eq_master1}) and (\ref{eq_master3}) are equivalent (note that this means that the IME framework is---just as the Redfield master equation---in general non-Lindbladian), the key advantage of Eq.~(\ref{eq_master3}) is that the $\Gamma_{mn,m'n'}(t)$ can be evaluated, for the system under consideration, exactly and, in general, systematically order-by-order. The formulation in terms of the jump operators $F_{mn}$, which can be obtained using Eq.~(\ref{eq_iq_eb}), not only provides a practical route for evaluating the integral in Eq.~(\ref{eq_master1}) but additionally provides, as we show below, a transparent framework for interpreting the fluorescence spectra. We emphasize that the IME is non-perturbative in $\Omega_p \text{~and~} \Omega_s$. 

It is instructive to connect the IME framework with other master equation frameworks, specifically a so-called laboratory-frame master equation framework and a so-called rotating-frame master equation framework. 
The terminology ``laboratory-frame master equation'' and ``rotating-frame master equation'' follows Ref.~\cite{shavit2019}, which treats a driven 2-level system. It is important to note that these terms refer to distinct approximations that are made when deriving the master equation and not to the actual frame in which the calculations are performed. In most cases, master equations are derived by employing some sort of rotating frame, even in the case of the laboratory-frame master equation. The labels ``laboratory-frame master equation'' and ``rotating-frame master equation'' allude to the stage of the derivation at which key approximations are made. The derivation of the laboratory-frame master equation follows the standard route~\cite{shavit2019}. A detailed derivation of the rotating-frame master equation for $N=3$ will be published elsewhere~\cite{BasakInPrep}.

For concreteness, we employ---as in Appendix~\ref{sec:appendA}---the rotating wave approximation ($\omega_p,\omega_s \gg \Omega_p, \Omega_s$). Employing a zero-temperature bath, we find that the IME, the rotating-frame master equation, and the laboratory-frame master equation for the three-level system can, if the jump operators are expressed in the invariant eigenbasis for all three master equation frameworks, be written in a unified way:
\begin{widetext}
\begin{align}
    &\dfrac{d}{dt}\rho_{q,R}(t) = -i\comm{H_{q,R}(t)}{\rho_{q,R}(t)}+ \nonumber\\&\qquad\qquad\smashoperator{\sum_{m,n,m',n'}}
    \left( \gamma_{\perp,mn,m'n'}\left[F_{m'n'}\rho_{q,R}(t)F_{mn}^{\dagger} -F_{mn}^{\dagger}F_{m'n'}\rho_{q,R}(t)\right]  + 
\gamma_{\parallel,mn,m'n'}\left[F_{m'n'}\rho_{q,R}(t)F_{mn}^{\dagger} -F_{mn}^{\dagger}F_{m'n'}\rho_{q,R}(t)\right] \right) + \text{H.c.} \,.
\label{eq_UME}
\end{align}
\end{widetext}
It should be noted that the laboratory-frame master equation is most commonly written in terms of the ``bare'' atomic levels $\ket{1}$, $\ket{2}$, and $\ket{3}$ and not in terms of the $\{|\mu_n\rangle\}$. 
For the example at hand, the invariant eigen basis coincides with the dressed-state basis, i.e., the eigen states of the system Hamiltonian in the rotating frame. This feature is used when transforming from the basis $\{ \ket{1}, \ket{2}, \ket{3} \}$ to the basis $\{|\mu_0\rangle,|\mu_+\rangle,|\mu_-\rangle\}$. 
The time independence of the dissipation coefficients $\gamma_{\perp,mn,m'n'}$ and $\gamma_{\parallel,mn,m'n'}$ in Eq.~(\ref{eq_UME}) is a consequence of making the rotating wave approximation when evaluating the dissipation coefficients [compare with $\Gamma_{mn,m'n'}(t)$ in Eq.~(\ref{eq_master3}); see Appendix~\ref{sec:appendC} for details].

The dissipation coefficients associated with the transverse and longitudinal decoherences, for the IME [see Appendix~\ref{sec:appendC}, Eqs.~(\ref{eq_decayrate_perp}) and (\ref{eq_decayrate_par})], rotating-frame (RF) master equation, and laboratory-frame (LF) master equation are
\begin{align}
    &\gamma_{\perp,mn,m'n'} \underset{\text{IME}}{=} \Gamma_{\perp}\xi^{\perp,12}_{mn}\xi^{\perp,12}_{m'n'}\left[\dfrac{(\kappa_{\perp}/2)^2}{(\kappa_{\perp}/2)^2 + (\alpha^{\perp,12}_{m'n'}-\omega_{\perp})^2}\right] \,,
    \label{eq_IMEperpdis}
    \\& \gamma_{\perp,mn,m'n'} \underset{\text{RF}}{=} \Gamma_{\perp}\xi^{\perp,12}_{mn}\xi^{\perp,12}_{m'n'}\left[\dfrac{(\kappa_{\perp}/2)^2}{(\kappa_{\perp}/2)^2 + (\alpha^{\perp,12}_{m'n'}-\omega_{\perp})^2}\right] \notag
     \\& \phantom{\gamma_{\perp,mn,m'n'} \underset{\text{RF}}{=}}\times (\delta_{m,m'}\delta_{n,n'}+\delta_{m,n}\delta_{m'n'}-\delta_{m,m'}\delta_{m,n}\delta_{n,n'})\,,
    \label{eq_RFperpdiss}
    \\&\gamma_{\perp,mn,m'n'} \underset{\text{LF}}{=} \Gamma_{\perp}\xi^{\perp,12}_{mn}\xi^{\perp,12}_{m'n'}\,,
    \label{eq_LFperpdiss}
\end{align}
    and
\begin{align}
    &\gamma_{\parallel,mn,m'n'} \underset{\text{IME}}{=} \Gamma_{\parallel}\xi^{\parallel}_{mn}\xi^{\parallel}_{m'n'}\left[\dfrac{(\kappa_{\parallel}/2)^2}{(\kappa_{\parallel}/2)^2 + (\alpha^{\parallel}_{m'n'}-\omega_{\parallel})^2}\right]\,, 
    \label{eq_IMEpardis}
    \\&\gamma_{\parallel,mn,m'n'} \underset{\text{RF}}{=} \Gamma_{\parallel}\xi^{\parallel}_{mn}\xi^{\parallel}_{m'n'}\left[\dfrac{(\kappa_{\parallel}/2)^2}{(\kappa_{\parallel}/2)^2 + (\alpha^{\parallel}_{m'n'}-\omega_{\parallel})^2}\right] \notag
    \\&\phantom{\gamma_{\parallel,mn,m'n'} \underset{\text{RF}}{=}}\times(\delta_{m,m'}\delta_{n,n'}+\delta_{m,n}\delta_{m'n'}-\delta_{m,m'}\delta_{m,n}\delta_{n,n'})\,,
    \label{eq_RFpardiss}
    \\&\gamma_{\parallel,mn,m'n'} \underset{\text{LF}}{=} \Gamma_{\parallel}\xi^{\parallel}_{mn}\xi^{\parallel}_{m'n'}\,.
    \label{eq_LFpardiss}
\end{align}
The quantities $\xi^{\perp,12}_{mn}$, $\xi^{\parallel}_{mn}$, $\alpha^{\perp,12}_{mn}$ and $\alpha^{\parallel}_{mn}$ are defined in Appendix~\ref{sec:appendB}.
As can be seen in Eqs.~(\ref{eq_RFperpdiss}) and (\ref{eq_RFpardiss}), the dissipation coefficients of the rotating-frame master equation can be expressed in terms of those of the IME. However, compared to the IME, the rotating-frame master equation neglects certain $mn,m'n'$ combinations (via the delta-functions). In contrast, Eqs.~(\ref{eq_LFperpdiss}) and (\ref{eq_LFpardiss}) show that the dissipation coefficients of the laboratory-frame master equation are obtained from the IME dissipation coefficients by neglecting their ``renormalization,'' i.e., by dropping the terms in the square brackets in Eqs.~(\ref{eq_IMEperpdis}) and (\ref{eq_IMEpardis}). 

The above discussion indicates that differences in physical observables obtained using the IME, the laboratory-frame master equation, and the rotating-frame master equation are expected to arise from the differences in the dissipation coefficients. Compared to the IME, the rotating-frame master equation neglects a subset of the dissipative terms, while the laboratory-frame master equation employs ``bare'' i.e., non-renormalized, dissipation coefficients. The resulting impact on the fluorescence spectra is analyzed in the next two sections.

As alluded to above, the IME is---as the Redfield master equation---non-Lindbladian and cannot, in general, be written in Lindbladian form. Inserting the dissipation coefficients for the laboratory-frame and rotating-frame master equations into Eq.~(\ref{eq_UME}), it can be readily shown that the laboratory-frame and rotating-frame master equations can be, as expected, rewritten in Lindbladian form. The advantage of the unified master equation, Eq.~(\ref{eq_UME}), is that it shows explicitly that the IME framework reduces to the laboratory-frame and rotating-frame master equations under appropriate simplifications. This indicates that the IME framework is more general and reduces to familiar formulations under appropriate assumptions.

Since the IME framework does not, in general, guarantee complete positivity, we need to devise a criterion that ensures that our results are physical. Rather than imposing additional secular approximations, which would allow us to force the IME to reduce to a master equation of Lindblad form, we retain the Redfield-level dissipative structure associated with the microscopic system-bath coupling and instead restrict ourselves to parameter combinations for which positivity is satisfied. This is done using a ``post-selection'' approach, in which we run a simulation and only report the results when positivity is satisfied at the level of our numerical precision. A systematic analysis into an {\em{a priori}} determination of positivity is beyond the scope of the current work.

\section{\label{sec:definition}Definition of Fluorescence Spectrum}
\subsection{Definition}
We use the unified master equation in Eq.~(\ref{eq_UME}), which incorporates the IME framework as well as the laboratory-frame and rotating-frame master equation frameworks, to calculate the driven qutrit's fluorescence spectrum, an experimentally accessible observable that probes the system’s internal dynamics during radiative relaxation~\cite{grove1977,ates2009,mollow1969,Glauber1963}. Focusing on the incoherent (inelastic) component, which dominates beyond saturation and carries nontrivial dynamical information, the spectrum in the rotating frame is defined as the Fourier transform of the first-order correlation function that is associated with the $\ket{2}\rightarrow \ket{1}$ transition~\cite{steck_quantum_atom_optics,lax1963,Chen2019}
\begin{align}
    S_{12}(\omega) = \dfrac{1}{2\pi}\lim_{t_0 \rightarrow \infty} \int\limits_{-\infty}^{\infty}d\tau e^{-i\omega \tau}\expval{\delta\sigma^{\dagger}_{12}(t_0)\delta\sigma^{\phantom{\dagger}}_{12}(t_0+\tau)}\,,
\end{align}
where $\delta \sigma_{12}(t)=\sigma_{12}(t)-\expval{ \sigma_{12}}_{\text{ss}}, \text{ with } \expval{ \sigma_{12}}_{\text{ss}}$ denoting the steady-state expectation value.
In addition to $S_{12}(\omega)$, we define the normalized fluorescence spectrum through
\begin{eqnarray}
S^N_{12}(\omega) = \frac{S_{12}(\omega)}{\text{max}\left[S_{12}(\omega)\right]}\,\vcenter{\hbox{.}}
\end{eqnarray}
Numerically, we obtain $S_{12}(\omega)$ using the quantum regression theorem (see Appendix~\ref{sec:appendD} for details).
For a coherently driven qubit, the fluorescence spectrum yields the well-known Mollow triplet, which arises from transitions between dressed states~\cite{ulrich2011,muller2007,boos2024,Stenquist2024,Ulhaq2013}. Section~\ref{sec:results} shows how the Mollow triplets for a three-level system are modified relative to those for a two-level system, using the IME as well as their limiting Lindbladian forms.

\subsection{Review of the Mollow triplet}
\label{sec:definition:review}
To set the stage for the discussion of the driven {\em{three-level}} system presented in Sec.~\ref{sec:results}, we review the emergence of the Mollow triplet for a relatively strongly driven {\em{two-level}} system. We discuss the spectrum in the rotating frame, where $\omega=0$ corresponds to the drive frequency $\omega_p$. For a strongly-driven two-level system with atomic states $\ket{1}$ and $\ket{2}$, we find that the rotating-frame master equation and the IME yield quite similar results. Forthcoming work~\cite{BasakInPrep} shows that the IME is equivalent to the generalized master equation framework considered in Ref.~\cite{shavit2019}, if the IME is truncated appropriately. In what follows, we explain the number of peaks, peak positions, and peak heights for a relatively strongly driven two-level system.

{\em{Number of peaks and peak positions:}} The peak positions of $S^N_{12}(\omega)$ and $S_{12}(\omega)$ for the strongly-driven two-level system can, to leading order, be obtained from the eigenenergies $E_+$ and $E_-$ of the dressed states $|\mu_+\rangle$ and $|\mu_-\rangle$, i.e., of the eigenstates of $H_{q,R}$. 
Since the bare atomic states $\ket{1}$ and $\ket{2}$ can be written as a superposition of $|\mu_+\rangle$ and $|\mu_-\rangle$, the three peaks of the Mollow triplet can be interpreted as corresponding to the transitions between the states $|\mu_+\rangle$ and $|\mu_-\rangle$ (side peaks at $\approx \pm \Omega_T$, where $\Omega_T = E_+-E_-$) and between the states $|\mu_+\rangle$ and $|\mu_+\rangle$ or between the states $|\mu_-\rangle$ and $|\mu_-\rangle$ (peak centered at $\omega\approx 0$). 

{\em{Peak heights:}} On resonance (i.e., for zero detuning), assuming a uniform bath spectrum, the height of the central peak is proportional to $1/(4\pi\Gamma_{\perp})$, whereas the height of the side peaks is proportional to $1/(12\pi \Gamma_{\perp})$; this shows that the central peak is higher.
For a large detuning, in contrast, the height of the central peak scales as $1/\Delta^6$, whereas the height of the side peaks scales as $1/\Delta^4$; this shows that the side peaks are higher.
The arguments just made explain the relative peak heights in the limits of small and large detuning but do not explain the asymmetry of the side peaks. 
To explain the difference in the heights of the side peaks, we also need to consider the populations and spectral function.
Since the central peak depends on the steady-state populations of both dressed states, the height of the central peak is directly proportional to the value of the spectral function at $\pm \Omega_T$. In contrast, the side peak at negative frequency, which is associated with the transition from $|\mu_+\rangle$ to $|\mu_-\rangle$, is proportional to the steady-state population of $|\mu_+\rangle$ and, correspondingly, to the spectral function at $\omega=\Omega_T$. Similarly, the side peak at positive frequency, which is associated with the transition from $|\mu_-\rangle$ to $|\mu_+\rangle$, is proportional to the steady-state population of $|\mu_-\rangle$ and, correspondingly, to the spectral function at $\omega=-\Omega_T$. This implies that the side peaks develop an asymmetry if the bath is structured (i.e., non-uniform). 

When the Rabi coupling strength decreases, results obtained using the rotating-frame master equation and the IME deviate.

\begin{figure*}[t]
    \centering
    \includegraphics[width=6.75in]{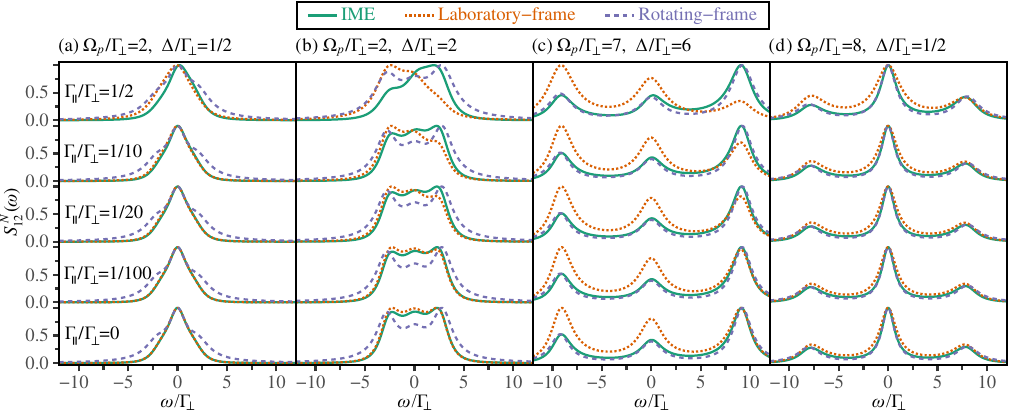}
    \caption{\label{fig:App1RS12}(color online) Application 1. Normalized fluorescence spectra $S^N_{12}(\omega)$ as a function of the angular frequency $\omega$, calculated using the IME (green solid lines), the laboratory-frame (orange dotted lines), and the rotating-frame (purple dashed lines), for (a) ($\Omega_p/\Gamma_{\perp}$, $\Delta/\Gamma_{\perp}$) =  $(2,\;0.5)$, (b) $(2,\;2)$, (c) $(7,\;6)$, and (d) $(8,\;0.5)$. Each panel shows, using equidistant vertical offsets,  spectra (from bottom to top) for $\Gamma_{\parallel}/\Gamma_{\perp} = 0$, $0.01$, $0.05$, $0.1$, $0.5$. The other parameters are: $\kappa_{\perp} = \kappa_{\parallel} = 30\Gamma_{\perp},\,\omega_{\perp} = \omega_{1}=3\times 10^5 \Gamma_{\perp},\,\omega_{\parallel} =0,\, \text{and }\omega_1-\omega_3 = 10\Gamma_{\perp}$. The bath temperature $T_b$ is set to zero.}
\end{figure*}
\begin{figure*}[t]
    \centering
    \includegraphics[width=6.75in]{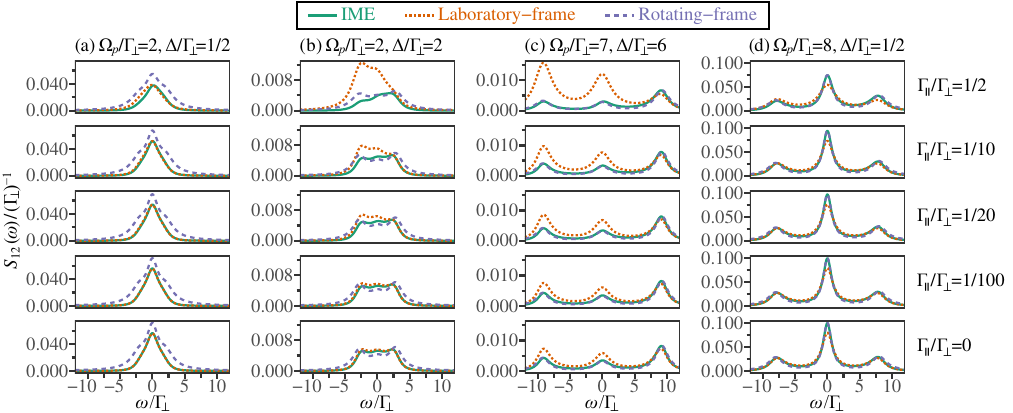}
    \caption{\label{fig:App1RS12_U}(color online) Application 1. Un-normalized fluorescence spectra $S_{12}(\omega)$ as a function of the angular frequency $\omega$ using the same parameter combinations and line conventions as those used in Fig.~\ref{fig:App1RS12}.}
\end{figure*}

\section{\label{sec:results}Fluorescence Spectra of a driven Qutrit}
\begin{figure*}[t]
    \centering
    \includegraphics[width=6.75in]{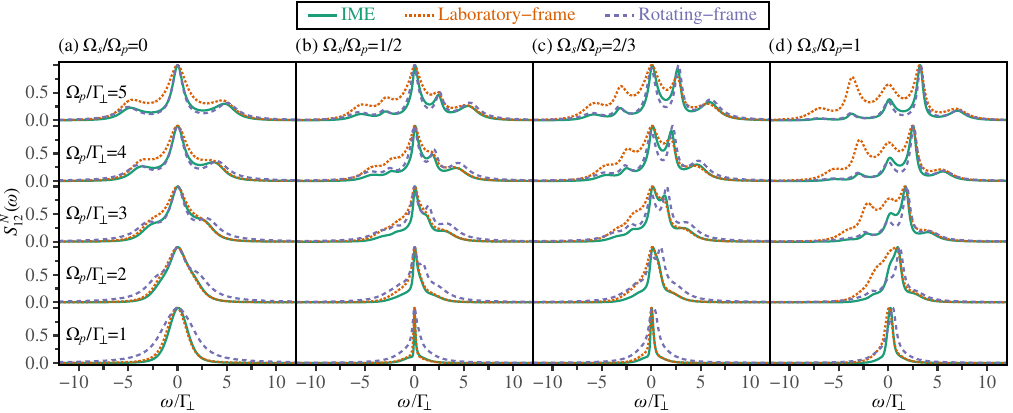}
    \caption{\label{fig:App2RS12}(color online) Application 2. Normalized fluorescence spectra $S^N_{12}(\omega)$ as a function of $\omega$, calculated using the IME (green solid lines), the laboratory-frame (orange dotted lines), and the rotating-frame (purple dashed lines), for (a) $\Omega_s/\Omega_p = 0$, (b) $1/2$, (c) $2/3$, and (d) $1$. Each panel shows, using equidistant vertical offsets,  spectra (from bottom to top) for $\Omega_p/\Gamma_{\perp} = 1$, $2$, $3$, $4$, $5$.  The other parameters are: $\Delta/\Gamma_{\perp} = 0.5,\,\kappa_{\perp}=12/\Gamma_{\perp},\,\kappa_{\parallel} = 40/\Gamma_{\perp},\,\Gamma_{\parallel}/\Gamma_{\perp} =0.1,\,\omega_{\perp} = \omega_{1}=3\times 10^5 \Gamma_{\perp},\,\omega_{\parallel} =0,\,\omega_1-\omega_3 = 10\Gamma_{\perp},\,\text{and } T_b =0$. 
   }
\end{figure*}

\begin{figure*}[t]
    \centering
    \includegraphics[width=6.75in]{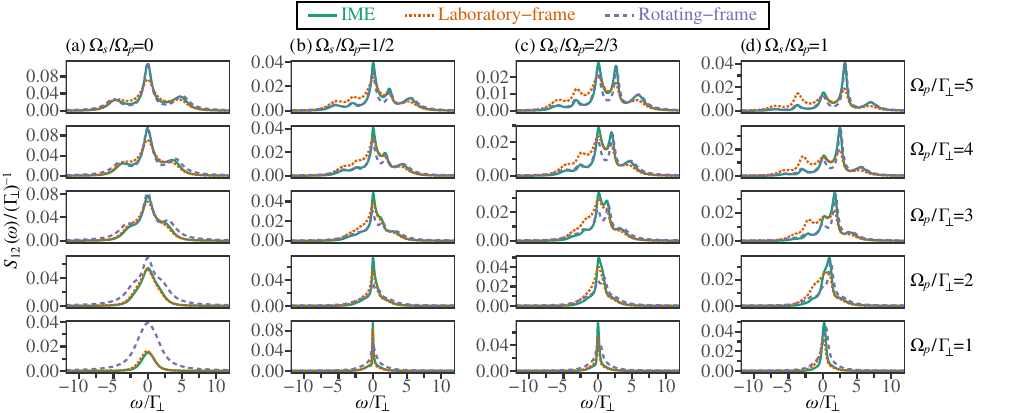}
    \caption{\label{fig:App2RS12_U}(color online) Application 2. Un-normalized fluorescence spectra $S_{12}(\omega)$ 
    as a function of the angular frequency $\omega$ using the same parameter combinations and line conventions as those used in Fig.~\ref{fig:App2RS12}.}
\end{figure*}

We now analyze the fluorescence spectra of the driven {\em{three-level}} system, or the qutrit, for Applications 1 (Figs.~\ref{fig:App1RS12} and \ref{fig:App1RS12_U}) and 2 (Figs.~\ref{fig:App2RS12} and \ref{fig:App2RS12_U}). Both the un-normalized spectra $S_{12}(\omega)$ and the normalized spectra $S^N_{12}(\omega)$ are considered. Compared to the spectra for the qubit, the spectra for the qutrit display modified line shapes and additional peaks. The resulting peak positions, widths, and relative weights encode the interplay of coherent driving and dissipation, providing a sensitive probe of system--bath effects such as broadening and spectral asymmetry. To understand the nuances of the fluorescence spectrum over the entire parameter regime, including the regime where the laboratory-frame master equation and the IME yield nearly identical results, it is useful to base the analysis on the eigenstates and eigenenergies of the Liouvillian generator ${\cal{L}}$ and not on the eigenstates and eigenenergies of $H_{q,R}$ as done in the review of the Mollow triplets presented in Sec.~\ref{sec:definition:review}. This can be understood intuitively by realizing that the dynamics of the density matrix is governed by the Liouvillian generator; an analysis of the system Hamiltonian is not sufficient.

Appendix~\ref{sec:appendE} shows, focusing on Application~1, that the key characteristics of the spectra shown in Figs.~\ref{fig:App1RS12} and \ref{fig:App1RS12_U} (number of peaks, peak positions, peak widths, and peak heights) can be explained by analyzing the eigenvalues and eigenvectors of the Liouvillian generator ${\cal{L}}$. Specifically, (i) the number of peaks reflects the number of distinct dynamical channels contributing to emission from state $\ket{2}$ to state $\ket{1}$, (ii) the peak positions are set by the characteristic oscillation frequencies associated with these modes, (iii) the peak widths are determined by their associated decoherence rates, and (iv) the peak heights are determined by how strongly the corresponding dynamical modes overlap with the operator that generates the fluorescence spectrum and by their contribution to the steady state density matrix.
Quite generically, it then follows that the differences in the spectra obtained using different master equations originate from what approximations are being made when deriving the dissipation coefficients (e.g., whether or not the drive is kept and which, if any, secular approximations are being made).
An important take-away message from our analysis is that a proper description of the features of the spectra of strongly-driven three-level systems in the presence of a structured bath requires a consistent open quantum system description, such as the one put forward in our work.


\subsection{Application 1: Qutrit with \texorpdfstring{$(1{\leftrightarrow}2)$}{1--2}- but without \texorpdfstring{$(2{\leftrightarrow}3)$}{2--3}-drive}
The system investigated here is shown in Fig.~\ref{fig:System}(a). In this case, the excited state $\ket{2}$ undergoes bath-induced dissipation via the transverse term of $H_I$. Since $\Omega_s$ is zero, the qutrit itself can be thought of as a driven qubit with coherent time-dependent $(1{\leftrightarrow}2)$-coupling (finite $\Omega_p$) whose ground state is coupled to state $\ket{3}$ via reservoir-induced dephasing (longitudinal term of $H_I$). While the driven qubit has been studied in detail within a generalized master equation framework (which is equivalent to the IME used in our work in certain limits~\cite{BasakInPrep}) and results have been carefully compared with those obtained using laboratory-frame and rotating-frame master equations~\cite{shavit2019}, we are not aware of extensions of the generalized framework to systems with $N>2$. Below, we show results for the IME framework [Eq.~(\ref{eq_master3})] and compare with results obtained using the laboratory-frame and rotating-frame master equations (while the derivation for the $N=3$ system for the latter is lengthy~\cite{BasakInPrep}, the steps follow standard procedures).

Figures~\ref{fig:App1RS12} and \ref{fig:App1RS12_U} show the normalized and un-normalized fluorescence spectra for four different $(\Omega_p/\Gamma_{\perp},\Delta/\Gamma_{\perp})$ combinations ($\Delta$ denotes the detuning, $\Delta=\omega_1-\omega_p$) and five different $\Gamma_{\parallel}/\Gamma_{\perp}$. In each panel, the spectra for different $\Gamma_{\parallel}/\Gamma_{\perp}$ are offset vertically. The parameters are chosen such that the IME spectra (green solid lines) are, for $\Gamma_{\parallel}/\Gamma_{\perp}=0$ (bottom-most set of spectra in each panel; this corresponds to a driven qubit), reproduced quite well by either the rotating-frame master equation [purple dashed lines; Figs.~\ref{fig:App1RS12}(c) and \ref{fig:App1RS12_U}(c)], the laboratory-frame master equation [orange dotted lines; Figs.~\ref{fig:App1RS12}(a) and \ref{fig:App1RS12_U}(a)], or both [Figs.~\ref{fig:App1RS12}(d) and \ref{fig:App1RS12_U}(d)]. This can be explained as follows: The dissipative terms in the laboratory-frame master equation are derived by neglecting the drive, which requires that the generalized Rabi frequency $\Omega_T$, $\Omega_T = (\Omega^2_p + \Delta^2)^{1/2}$, be much smaller than the transverse spectral width $\kappa_{\perp}$ of the bath. The dissipative terms in the rotating-frame master equation, in contrast, are derived by neglecting drive-induced oscillatory terms, which requires that $\Gamma_{\perp}$ be much smaller than $\Omega_T$. 
For both Figs.~\ref{fig:App1RS12} and \ref{fig:App1RS12_U}, panels (a)--(d) correspond to $(\Omega_T/\kappa_{\perp},\,\Gamma_{\perp}/\Omega_T)=(0.0687,\,0.485),\; (0.0943,\,0.354),\; (0.307,\,0.108),$ and $(0.267,\,0.125)$, respectively. This explains why the $\Gamma_{\parallel}=0$ spectra obtained using the IME are reproduced quite well by frameworks of more limited applicability.

Figures~\ref{fig:App1RS12} and \ref{fig:App1RS12_U} show that the agreement between the results within the different frameworks deteriorates as $\Gamma_{\parallel}/\Gamma_{\perp}$ increases for the $(\Omega_p/\Gamma_{\perp},\,\Delta/\Gamma_{\perp})$ parameters considered. This deterioration is schematically shown in the right part of Fig.~\ref{fig:System}(a), which illustrates that the validity regimes of the laboratory-frame and rotating-frame master equations (red and blue rectangles) decrease with increasing $\Gamma_{\parallel}/\Gamma_{\perp}$; note, though, that the level of deterioration depends on the specific parameters. 
Figures~\ref{fig:App1RS12}(a), \ref{fig:App1RS12}(b), \ref{fig:App1RS12_U}(a), and \ref{fig:App1RS12_U}(b) also show that a finite $\Gamma_{\parallel}$ can introduce an asymmetry, with bias in the IME spectra toward positive frequencies, that is absent for $\Gamma_{\parallel}/\Gamma_{\perp}=0$ and not properly captured by either the laboratory-frame and rotating-frame master equations as $\Gamma_{\parallel}/\Gamma_{\perp}$ increases. 

Inspection of the un-normalized spectra, Fig.~\ref{fig:App1RS12_U}, shows that the peak heights of the spectra can deviate by more than a factor of two for spectra calculated by different master equation approaches. In the top panel of Fig.~\ref{fig:App1RS12_U}(b), e.g., the peak height at $\omega/\Gamma_{\perp} \approx -2.5$ differs appreciably for the three master equation frameworks considered. These differences should be measurable with state-of-the-art experimental set-ups. Note also that the un-normalized spectra are scaled by $(\Gamma_{\perp})^{-1}$. This implies that the dimensionful spectra can be quite large, facilitating experimental observation. 

\subsection{Application 2: Qutrit embedded in cavity and coupled to bath}
The system investigated here is shown in Fig.~\ref{fig:System}(b). We consider an experimentally realizable quantum-dot--cavity system~\cite{mucke2010,mi2011,muller2007,ulrich2011} in which the $(1{\leftrightarrow}2)$- and $(2{\leftrightarrow}3)$-transitions of the quantum dot are driven (we use $\Delta=\omega_1-\omega_p=\omega_3-\omega_s$). Focusing on the regime in which the quantum dot decoheres predominantly through transverse coupling to a leaky cavity that itself interacts strongly with a thermal bath, the single-mode cavity effectively acts as a structured reservoir. The strong cavity--bath coupling leads to a broadening of the delta-function-like cavity resonance to a Lorentzian spectral density that enables dissipation over a finite frequency window. The resulting qutrit--cavity interaction is predominantly incoherent and the cavity degrees of freedom can be treated as stationary on timescales relevant for the qutrit dynamics, thereby justifying an effective ``cavity-as-bath'' description. It follows that the interaction Hamiltonian $H_I$, Eq.~(\ref{eq_intham}), applies to the quantum-dot--cavity system if the replacement $b \rightarrow c$ (with $c$ denoting a cavity operator) is made in the transverse coupling term (see Appendix~\ref{sec:appendB}). In this setting, energy relaxation (transverse coupling) and pure dephasing (longitudinal coupling) are cavity- and thermal-bath induced, respectively. 

Figures~\ref{fig:App2RS12} and \ref{fig:App2RS12_U} show the normalized and un-normalized fluorescence spectra using parameters realizable in state-of-the-art quantum-dot--cavity platforms: transition, drive, and cavity frequencies around $300$ THz; coupling strength $g \sim 2-6$ GHz; cavity decay rate $\kappa_{c} \sim 6-20$ GHz, which sets the spectral width $\kappa_{\perp}$ for the transverse coupling; maximum dephasing rate $\Gamma_{\parallel} \sim 0.1$ GHz; and drive strengths $\lesssim 50$ GHz~\cite{mucke2010,mi2011,muller2007,ulrich2011}. Figures~\ref{fig:App2RS12} and \ref{fig:App2RS12_U} consider four different coupling ratios $\Omega_s/\Omega_p$ [for $\Omega_s=0$, Figs.~\ref{fig:App2RS12}(a) and \ref{fig:App2RS12_U}(a), Application 2 coincides with Application 1].
In each panel, the spectra corresponding to different normalized probe drive strengths $\Omega_p/\Gamma_{\perp}$ are vertically offset. The IME results (green solid lines) for $\Omega_p/\Gamma_{\perp}=1$ (bottom-most set of spectra in each panel) are quite well captured by the laboratory-frame approach and those for $\Omega_p/\Gamma_{\perp}=5$ (top-most set of spectra) are quite well captured by the rotating-frame approach. With increasing $\Omega_p/\Gamma_{\perp}$, the system transitions from a regime where the laboratory-frame approach is valid, to an intermediate regime where both the laboratory- and rotating-frame approaches fail, to a regime where the rotating-frame approach is valid. 

Comparing the finite-$\Omega_s$ spectra in Figs.~\ref{fig:App2RS12}(b)-\ref{fig:App2RS12}(d) and \ref{fig:App2RS12_U}(b)-\ref{fig:App2RS12_U}(d) with the corresponding $\Omega_s=0$ spectra in Figs.~\ref{fig:App2RS12}(a) and \ref{fig:App2RS12_U}(a), it is evident from the emergence of additional peaks and of spectrum-asymmetry with respect to $\omega=0$ that the $(2{\leftrightarrow}3)$-drive introduces one or more new time scales. Our analysis shows that the new spectral peaks appear because all three bare states contribute to the dressed eigenstates for $\Omega_s \ne 0$. Consequently, transitions that involve state $\ket{1}$ probe all three dressed states for finite $\Omega_s$ and not just, as for $\Omega_s=0$, two dressed states.
As the probe drive strength increases (moving upward within each panel), the spectrum evolves from featuring a single dominant peak to featuring a broadened plateau to featuring a multi-peak structure.
While the laboratory- and rotating-frame approaches capture some of these intricate spectral features, neither provides a quantitatively correct description for $\Omega_p/\Gamma_{\perp}\approx2-4 \text{~and~} \Omega_s\neq 0$. 

\section{\label{sec:summary}Summary and Outlook}
Our work develops much needed theoretical tools for treating open, time-dependent quantum systems whose dynamics is governed by disparate timescales. Although the approach accommodates arbitrary time-dependent driving, for which Floquet theory may be inapplicable, and allows for systematic order-by-order improvements, the examples considered here focus on periodically driven systems. Specifically, we consider systems that admit a time-independent rotating-frame Hamiltonian under the rotating-wave approximation, allowing us to clearly demonstrate how drive-induced dressing modifies the dissipative dynamics without requiring a perturbative treatment of the drive strength. 

First, we showed that the invariant-based master equation (IME) is applicable broadly, beyond prototypical systems such as the harmonic oscillator and two-level system~\cite{boubakour2025,wu2022,dann2018,dann2019}, and not limited to perturbative drive strengths. 
Second, we showed that the IME provides a unifying framework that reduces, under appropriate assumptions, to the frequently used laboratory-frame master equation framework and the rotating-frame master equation framework, which are applicable in ``opposing'' limiting regimes. 
Third, we showed that the simplest extension of a driven two-level system, namely a driven three-level system, exhibits new physics that is not captured by other approaches, thereby underlining the need for novel master equation approaches.
Spectral asymmetry, peak shifts, and incorrect steady states were identified as clear signatures of the breakdown of conventional treatments. We established that these failures originate in incorrect frequency sampling of bath spectral functions or in neglecting contributions, as a consequence of the secular approximation, in standard approaches. The IME framework consistently captures drive-induced dressed-state transitions. Driven open qutrits serve as versatile platforms for quantum thermal machines~\cite{Mohan2025,Cangemi2024}, quantum-memory protection~\cite{Zhao2025}, and quantum synchronization~\cite{Jaseem2020}. The IME therefore provides a natural framework to investigate how arbitrary driving and structured reservoirs jointly govern the performance of quantum technologies.

Looking ahead, it will be critical to confront our IME predictions with experiment and to extend the framework to $N$-level systems with $N>3$ and other quantum hybrid platforms. The IME is, e.g., well suited to treat a range of cavity-QED and solid-state platforms that feature structured reservoirs and tunable spectral widths. Another extension of our work is to non-periodic, time-dependent driving protocols---such as pulses, ramps, and shortcuts to equilibration---directly relevant to quantum control and state-preparation tasks in systems with competing dissipative channels. Last, an important future direction is the generalization to non-Markovian regimes.

\begin{acknowledgments}
We thank M.~Boubakour, T.~Busch, and T.~Fogarty for insightful discussions at the early stage of this project; we also thank X. Molenda for discussions. 
This work is supported by the W.~M.~Keck Foundation. This material is also based upon work supported by the Air Force Office of Scientific Research under Award No.~FA9550-24-1-0106. A.~J.~acknowledges funding from the National Science Foundation under award number 2441706.
The computing for this project was performed at the OU Supercomputing Center for Education \& Research (OSCER) at the University of Oklahoma (OU).
\end{acknowledgments}

\section*{Data availability}
Some of the data that support the findings of this article are openly available~\cite{datarepo}, embargo periods may apply. Other are not publicly available upon publication because it is not technically feasible and/or the cost of preparing, depositing, and hosting the data would be prohibitive within the terms of this research project. The data are available from the authors upon reasonable request.

\appendix
\section{\label{sec:appendA} Invariant for the Qutrit System}
This appendix derives the dynamical invariant for a qutrit system and the associated phases, building on earlier two-level~\cite{xichen2011} and three-level studies~\cite{jin2025}. The qutrit Hamiltonian $H_q(t)$ in the laboratory frame is given by
\begin{align}
H_q(t)
&= -\omega_1 \ket{1}\bra{1} - \omega_3 \ket{3}\bra{3}
\nonumber\\
&\quad + \frac{1}{2}\Omega_p
\left(e^{i\omega_p t}+e^{-i\omega_p t}\right)
\left(\ket{1}\bra{2}+\ket{2}\bra{1}\right)
\nonumber\\
&\quad + \frac{1}{2}\Omega_s
\left(e^{i\omega_s t}+e^{-i\omega_s t}\right)
\left(\ket{3}\bra{2}+\ket{2}\bra{3}\right)\,.
\label{eq_qutrit}
\end{align}
The time dependence of $H_q(t)$ is handled by moving to a rotating frame and employing the rotating wave approximation (RWA)~\cite{Paing2025}. Defining the rotation operator $U_R(t)$,
\begin{align}
U_R(t) &= e^{-iDt}, \quad
D = \omega_p\ket{1}\bra{1}
  + \omega_s\ket{3}\bra{3}\,,
\label{eq_rot_op}
\end{align}
the qutrit Hamiltonian $H_{q,R}$ in the rotating frame is given by
\begin{align}
H_{q,R}
&= [U_R(t)]^{\dagger} H_q(t) U_R(t)
   - i [U_R(t)]^{\dagger} \frac{d}{dt} U_R(t)\,,\phantom{SP\;}
\label{eq_rot_gen}
\end{align}
\begin{align}
H_{q,R}
= -\Delta_p \ket{1}\bra{1}
   - \Delta_s \ket{3}\bra{3}
& + \frac{1}{2}\Omega_p
\left(\ket{2}\bra{1}+\ket{1}\bra{2}\right)
\nonumber\\
& + \frac{1}{2}\Omega_s
\left(\ket{2}\bra{3}+\ket{3}\bra{2}\right)\,,
\label{eq_qutrit_rot}
\end{align}
where the detunings are defined as  $\Delta_p=\omega_2-\omega_p$ and $\Delta_s=\omega_3-\omega_s$. Throughout, 
we work with equal detunings ($\Delta=\Delta_p=\Delta_s$), i.e., we set
$\omega_s=\omega_3-\omega_2+\omega_p$. 
The RWA, which neglects terms proportional to $e^{\pm 2i \omega_p t}$ and $e^{\pm 2i \omega_s t}$, is valid for drive frequencies $\omega_p$ and $\omega_s$ much larger than the Rabi coupling strengths $\Omega_p$ and $\Omega_s$. This is justified 
for quantum dot systems, such as 
those considered in Application~2, where
the transition and driving frequencies are $\sim300$ THz, whereas the Rabi couplings are $<50$ GHz~\cite{mucke2010,mi2011,muller2007,ulrich2011}.
We emphasize that the Hamiltonian $H_{q,R}$, obtained by employing the RWA, is time independent for the scenarios considered in this work.

To construct an invariant $I_q(t)$ that satisfies Eq.~(\ref{eq_def_iq}), the boundary condition $\comm{I_q(0)}{H_{q,R}}=0$ is imposed at time $t=0$. This condition ensures that $I_q(0)$ and $H_{q,R}$ have common eigenvectors. Guided by the known structure of the eigenvectors at $t=0$, we parametrize the eigenvectors of the invariant in terms of the unknown angles $\phi(t)$, $\zeta(t)$ and $\eta(t)$,
\begin{align}
\ket{\mu_0(t)} 
&= \cos\bm{(}\zeta(t)\bm{)} \ket{1} - \sin\bm{(}\zeta(t)\bm{)} \ket{3}\,, \nonumber
\\
\ket{\mu_+(t)}
&= \sin\bm{(}\phi(t)\bm{)} \sin\bm{(}\zeta(t)\bm{)} \ket{1} 
 + \sin \bm{(}\phi(t)\bm{)}\cos\bm{(}\zeta(t)\bm{)} \ket{3} \nonumber \\ 
&\phantom{= \sin\bm{(}\phi(t)\bm{)} \sin\bm{(}\zeta(t)\bm{)} \ket{1}\;}
 + e^{-i\eta(t)}\cos\bm{(}\phi(t)\bm{)} \ket{2}\,, \nonumber \\
\ket{\mu_-(t)} 
&= \cos\bm{(}\phi(t)\bm{)} \sin\bm{(}\zeta(t)\bm{)} \ket{1} 
 + \cos \bm{(}\phi(t)\bm{)}\cos\bm{(}\zeta(t)\bm{)} \ket{3} \nonumber \\
&\phantom{= \cos\bm{(}\phi(t)\bm{)} \sin\bm{(}\zeta(t)\bm{)} \ket{1}\;}
 -e^{-i\eta(t)} \sin\bm{(}\phi(t)\bm{)} \ket{2}\,,
\end{align}
with the corresponding eigenvalues $\lambda_0=0$, $\lambda_+=1$, and $\lambda_-=-1$. The boundary conditions set the initial values of the angles:
\begin{align}
\phi(0)
  &= \left[\tan^{-1}\left(\sqrt{\Omega^2_s+\Omega^2_p}/\Delta\right)\right]/2\,, \nonumber\\
\eta(0)
  &= 0\,, \nonumber\\
\zeta(0)
  &= \tan^{-1}\left(\Omega_p / \Omega_s\right)\,.
\end{align}
Using
\begin{align}
I_q(t) 
= \textstyle\sum\limits_n \lambda_n |\mu_n(t)\rangle\langle\mu_n(t)|\,,
\end{align} 
we find
\begin{align}
I_q(t)
= &\cos \bm{\bigl(}2\phi\left(t\right)\bm{\bigr)}\left[\ket{2}\bra{2} -\cos^2\bm{\bigl(}\zeta\left(t\right)\bm{\bigr)}\ket{3}\bra{3}-\sin^2\bm{\bigl(}\zeta\left(t\right)\bm{\bigr)}\ket{1}\bra{1}  \right]
\nonumber\\
& + \sin\bm{\bigl(}2\phi\left(t\right)\bm{\bigr)} \cos\bm{\bigl(}\zeta\left(t\right)\bm{\bigr)}  \left[e^{-i\eta\left(t\right)}\ket{2}\bra{3} + e^{i\eta\left(t\right)}\ket{3}\bra{2}\right]
\nonumber\\
&+  \sin \bm{\bigl(}2\phi\left(t\right)\bm{\bigr)} \sin\bm{\bigl(}\zeta\left(t\right)\bm{\bigr)}  \left[e^{-i\eta\left(t\right)}\ket{2}\bra{1} + e^{i\eta\left(t\right)}\ket{1}\bra{2}\right]
\nonumber\\
&- \cos \bm{\bigl(}2\phi\left(t\right)\bm{\bigr)}\sin\bm{\bigl(}\zeta\left(t\right)\bm{\bigr)}\cos\bm{\bigl(}\zeta\left(t\right)\bm{\bigr)} \left(\ket{3}\bra{1} + \ket{1}\bra{3}\right)\,.
\end{align}
Substituting the invariant and the Hamiltonian into the first part of Eq.~(\ref{eq_def_iq}), the equations of motion for the parameters $\phi(t)$, $\zeta(t)$, and $\eta(t)$ read
\begin{align}
\dot{\phi}\left(t\right) 
&= -\dfrac{1}{2}\sqrt{\Omega_s^2 + \Omega_p^2} \sin\bm{\bigl(}\eta\left(t\right)\bm{\bigl)}\,, \nonumber 
\\
\dot{\eta}\left(t\right) 
&= \Delta-\sqrt{\Omega_s^2 + \Omega_p^2} \dfrac{\cos\bm{\bigl(}\eta\left(t\right)\bm{\bigl)}}{\tan\bm{\bigl(}2\phi\left(t\right)\bm{\bigl)}}\,\vcenter{\hbox{,}} \text{ and} \nonumber
\\ 
\dot{\zeta}\left(t\right) 
&= 0\,.
\end{align}
By inspection, it follows that $\dot{\eta}(t) = 0$ and $\dot{\phi}(t) = 0$, which implies $\eta = \eta(0) = 0$, $\phi = \phi(0) = \left[\tan^{-1}\left(\sqrt{\Omega^2_s+\Omega^2_p}/\Delta\right)\right]/2$, and $\zeta = \zeta(0) = \tan^{-1}\left(\Omega_p / \Omega_s\right)$. Notably, $\phi$, $\zeta$, and $\eta$ are independent of time. We emphasize that this follows from the time independence of the Hamiltonian $H_{q,R}$. Correspondingly, the invariant $I_q$ and its eigenvectors $|\mu_n\rangle$, where $n$ stands for $0$, $+$, or $-$, are also time-independent. 

The Lewis-Riesenfeld phases $\alpha_n(t)$ are, in general, determined by the equation [see Eq.~(\ref{eq_def_iq})] 
\begin{align}
\dot{\alpha}_n (t) 
= \expval{i\dfrac{\partial }{\partial t} -H_{q,R}}{\mu_n(t)}\,.
\label{eq_lrp}
\end{align}
Inserting the results from above, we find
\begin{align}
\dot{\alpha}_0 
= \expval{i\dfrac{\partial }{\partial t} -H_{q,R}}{\mu_0} = \Delta
\end{align} 
and
\begin{align}
\dot{\alpha}_{\pm} 
= \expval{i\dfrac{\partial }{\partial t} -H_{q,R}}{\mu_{\pm}} 
= \dfrac{\Delta}{2}\mp \dfrac{1}{2}\sqrt{\Delta^2 + \Omega^2_s+\Omega^2_p}\,.
\end{align}
Since the $\dot{\alpha}_n$ are independent of time, we find 
$\alpha_n(t) = \dot{\alpha}_n t$.

Invariants for three-level systems have been previously derived in the context of shortcuts to adiabaticity and quantum control protocols~\cite{Li2021,Zhao2004,Kang2017,jin2025}. Our work builds on these formulations and derives the invariant for a qutrit system that is time independent in the rotating frame, yielding phases $\alpha_n(t)$ that vary linearly with time.

\section{\label{sec:appendB}Invariant-based Master Equation for Application 2}
Using the Redfield master equation, this appendix 
 presents the microscopic derivation of the IME for a driven qutrit coupled to a leaky cavity and an additional reservoir, which is used to obtain the green solid lines presented in Figs.~\ref{fig:App2RS12} and \ref{fig:App2RS12_U}. While this derivation is carried out explicitly for the system studied in Application~2, it builds on
IME
formulations developed in the literature for driven open quantum systems~\cite{wu2022,boubakour2025,dann2019,dann2018}. The resulting formalism can be directly adapted to Application~1 by replacing the cavity operators with bath operators and setting 
the $(2{\leftrightarrow}3)$-drive to zero.

We start in the laboratory frame. The total Hamiltonian $H(t)$ includes the qutrit Hamiltonian $H_q(t)$ [see Eq.~(\ref{eq_qutrit})], which is coupled to a cavity and a thermal bath,
\begin{equation}
    H(t) = H_q(t) + H_c+ H_b+ H_{qc}+ H_{qb} + H_{cb}\,,
\end{equation}
where 
the Hamiltonian $H_c$ of the single-mode cavity with cavity frequency $\omega_c$ is given by
\begin{equation}
\label{eq_cavity}
    H_c = \omega_c c^{\dagger}c^{\vphantom{\dagger}}\,,
\end{equation}
and the bath Hamiltonian $H_b$ is given by Eq.~(\ref{eq_ham_bath}).
In Eqs.~(\ref{eq_cavity}) and (\ref{eq_ham_bath}), $c^{\dagger}$ and $b_k^{\dagger}$ are bosonic operators that excite the cavity mode and the $ k$-th bath mode, respectively.
The total Hamiltonian contains three coupling terms: the qutrit-cavity coupling term $H_{qc}$, the qutrit-bath coupling term $H_{qb}$, and the cavity--bath coupling term $H_{cb}$. The presence of the latter allows us, as shown below, to treat the cavity as a second bath and to derive a master equation whose coherent dynamics is governed by the qutrit degrees of freedom.
The coupling $H_{qc}$, 
\begin{equation}
    H_{qc} = g\left(\sigma^{\vphantom{\dagger}}_{12}+\sigma^{\dagger}_{12}\right)\left(c^{\dagger}+c^{\vphantom{\dagger}}\right) \mbox{ with } \sigma_{12} = \ket{1}\bra{2}\,,
    \label{eq_Hqc}
    \end{equation}
     induces transitions between qutrit levels $\ket{1}$ and $\ket{2}$.
    The coupling $H_{qb}$,  
    \begin{equation}
         H_{qb} = \textstyle\sum\limits_k g_{\parallel,k}\left(\sigma_{33}-\sigma_{11}\right)\left(b^{\dagger}_k+b^{\vphantom{\dagger}}_k\right) \mbox{ with } \sigma_{nn} = \ket{n}\bra{n}\,,
         \label{eq_Hqb}
\end{equation}
accounts for bath induced dephasing between qutrit levels $\ket{1}$ and $\ket{3}$. 
This form of the system--bath interaction Hamiltonian is motivated by what is observed experimentally in setups of quantum dots coupled to a cavity~\cite{mucke2010,mi2011,muller2007,ulrich2011,maisch2024}. The treatment of more general qutrit-bath Hamiltonians is relegated to future work. 
The cavity--bath interaction Hamiltonian is given by
\begin{align}
    H_{cb} = \textstyle\sum\limits_k g_{cb,k} \left(c^{\dagger}+c^{\vphantom{\dagger}}\right)\left(b^{\dagger}_k+b^{\vphantom{\dagger}}_k\right)\,.
    \label{eq_Hcb}
\end{align}

As described in  Appendix~\ref{sec:appendA}, to render the qutrit Hamiltonian time-independent, we switch to the rotating frame using the rotation operator $U_R(t)$ given in Eq.~(\ref{eq_rot_op}). The total Hamiltonian $H_R(t)$ in the rotating frame is obtained from Eq.~(\ref{eq_rot_gen}) by replacing $H_q(t)$ with the total Hamiltonian $H(t)$. This transformation alters 
the qutrit Hamiltonian $H_q(t)$: the rotated qutrit Hamiltonian $H_{q,R}$ is given in Eq.~(\ref{eq_qutrit_rot}). The qutrit-cavity interaction Hamiltonian $H_{qc,R}(t)$ in the rotating frame is given by
\begin{align}
    H_{qc,R}(t) = g\left(e^{-i\omega_p t}\sigma^{\vphantom{\dagger}}_{12}+e^{i\omega_p t}\sigma^{\dagger}_{12}\right)\left(c^{\dagger}+c^{\vphantom{\dagger}}\right)\,.
\end{align}
The qutrit-bath interaction is unaltered when moving to the rotating frame since $\sigma_{11}$ and $\sigma_{33}$ contain products of lowering and raising operators. It follows: $H_{qb,R} = H_{qb}$. The other terms in the Hamiltonian, namely, $H_c$, $H_b$, and $H_{cb}$, are also unaffected by the rotation since $U_R(t)$ and $[U_R(t)]^{\dagger}$ only act on the qutrit's Hilbert space, i.e., $H_{c,R} = H_{c}$, $H_{b,R} = H_{b}$, and $H_{cb,R} = H_{cb}$.

In the rotating frame, the composite-system density matrix $\rho_R(t)$ evolution is governed by the von Neumann equation
\begin{equation}
    \dfrac{d}{dt}\rho_R(t) = -i\comm{H_R(t)}{\rho_R(t)}\,.
\end{equation}
As is 
common 
in the derivation of a master equation, we next move to the interaction picture by using the evolution operator $U(t)$,
\begin{align}
    U(t) &= U_q(t)U_c(t)U_b(t) \,, \label{eq_21_Ut_21}
    \\
    U_c(t) &= e^{-iH_c t} = e^{-i\omega_cc^{\dagger}c^{\vphantom{\dagger}}t} \,, \label{eq_23_Uc_23}
     \\
     U_b(t) &= e^{-iH_b t} = e^{-i\sum_k \omega_{b,k} b^{\dagger}_kb^{\vphantom{\dagger}}_k t} = \textstyle\prod\limits_k e^{-i\omega_{b,k}b^{\dagger}_kb^{\vphantom{\dagger}}_kt}\,, \label{eq_24_Ub_24}
\end{align}
where the evolution operator $U_q(t)$ of the qutrit, which is given in Eq.~(\ref{eq_propagator}), is expressed in terms of the eigenstates of the invariant. As discussed in Appendix~\ref{sec:appendA}, the time independence of the qutrit Hamiltonian results in the time independence of the invariant eigenstates. 
As a consequence, $U_q(t)$ in Eq.~(\ref{eq_propagator}) reduces to
\begin{align}
    U_q(t) = \textstyle\sum\limits_{m} e^{i\dot{\alpha}_mt}\ket{\mu_m}\bra{\mu_m}\,.
    \label{eq_Uq}
\end{align}
In the interaction picture, the von-Neumann equation reads
\begin{align}
     \dfrac{d}{dt}\tilde{\rho}(t) &= -i\comm{\tilde{H}(t)}{\tilde{\rho}(t)}\,,\label{eq_vN}
     \\\tilde{\rho}(t) &= U^{\dagger}(t) \rho_R(t) U(t)\,,
     \\\tilde{H}(t) &= U^{\dagger}(t) H_R(t) U(t)-iU^{\dagger}(t)\dfrac{dU(t)}{dt} \nonumber \\&= \tilde{H}_{qc}(t) + \tilde{H}_{qb}(t) + \tilde{H}_{cb}(t)\,, \\
     \tilde{H}_{qc}(t) &= U^{\dagger}(t) H_{qc,R} U(t)\,, \nonumber \\
     \tilde{H}_{qb}(t) &= U^{\dagger}(t) H_{qb,R} U(t)\,, \nonumber \\
     \tilde{H}_{cb}(t) &= U^{\dagger}(t) H_{cb,R} U(t)\,.
\end{align}

To construct the Hamiltonian in the interaction picture, the operators that appear in the Hamiltonian need to be transformed to the interaction picture. 
Let $A(t)$ and $\tilde{A}(t)$ be system operators in the Schr\"odinger and interaction picture, respectively [for now, $A(t)$ is unspecified; in our case, we have to consider---as discussed below---the system operators $A_{qb}$ and $A_{qc}(t)$]. The transformation between the two pictures can be obtained using Eq.~(\ref{eq_21_Ut_21}) and Eqs.~(\ref{eq_23_Uc_23})--(\ref{eq_Uq}),
\begin{align}
\tilde{A}(t)
&= U^{\dagger}(t) A(t) U(t) \nonumber\\
&= U^{\dagger}_q(t) A U_q(t) \nonumber\\
&= \textstyle\sum\limits_{m,n}
   e^{-i\dot{\alpha}_m t}
   \ket{\mu_m}\mel{\mu_m}{A(t)}{\mu_n}\bra{\mu_n}e^{i\dot{\alpha}_n t}
   \nonumber\\
&= \textstyle\sum\limits_{m,n}
   e^{i(\dot{\alpha}_n-\dot{\alpha}_m)t}
   \mel{\mu_m}{A(t)}{\mu_n}
   \ket{\mu_m}\bra{\mu_n}\,.
\end{align}
Since the qutrit Hamiltonian is time independent for the driving terms considered in our work, the invariant and its eigenstates are also time independent. As a result, the amplitudes of the matrix elements remain time independent and the time dependence enters only through phases that grow linearly with time [see the $(\dot{\alpha}_n-\dot{\alpha}_m)t$ terms]. These phases arise from two sources: the Lewis-Riesenfeld phases, which are proportional to the eigenenergies times the time $t$, and the ``explicit rotating-frame phases'' of the system operators due to the drive, which are likewise proportional to $t$. In this situation, the formal driving timescale $\tau_D$ \cite{dann2018}, defined through the time curvature of these phases, is 
infinite. To proceed, the matrix elements are decomposed into a time-independent amplitude, a constant phase $\varsigma^A_0$, and an explicitly time-dependent phase $\varsigma^A(t)$,
\begin{align}
    \mel{\mu_m}{A(t)}{\mu_n} = \abs{\mel{\mu_m}{A(t)}{\mu_n}}e^{i\varsigma^A_0 + i\varsigma^A(t)} = \xi^{A}_{mn} e^{i\varsigma^A(t)}\,,
    \label{eq_mel}
\end{align}
where 
\begin{align}
\xi^{A}_{mn} =\abs{\mel{\mu_m}{A(t)}{\mu_n}}e^{i\varsigma^A_0 }.
\end{align} 
Using the jump operators $F_{mn}$ defined in Eq.~(\ref{eq_jop}),
the system operator $\tilde{A}(t)$ in the interaction picture becomes [using Eqs.~(\ref{eq_mel}) and (\ref{eq_jop})]
\begin{equation}
    \tilde{A}(t) = \sum_{m,n} e^{i(\dot{\alpha}_n-\dot{\alpha}_m)t}e^{i\varsigma^A(t)} \xi^{A}_{mn} F_{mn}= \sum_{m,n} e^{i\theta^A_{mn}(t)}\xi^{A}_{mn} F_{mn}\,,
\end{equation}
where $\dot{\alpha}_n$ is defined in Eq.~(\ref{eq_lrp}) and $\theta^A_{mn}(t)$ contains contributions from the 
Lewis-Riesenfeld phases and the time-dependent phase $\varsigma^A(t)$,
\begin{eqnarray}
\theta^A_{mn}(t)=(\dot{\alpha}_n-\dot{\alpha}_m)t+\varsigma^A(t)\,.
\label{eq_thetaphase}
\end{eqnarray}
If the system operator $A(t)$ is Hermitian, one can write
\begin{flalign}
\label{eq_appendixB}
   \tilde{A}(t) = \sum_{m,n} e^{i\theta^A_{mn}(t)}\xi^{A}_{mn} F_{mn} =  \sum_{m',n'} e^{-i\theta^A_{m'n'}(t)}(\xi^{A}_{m'n'})^* F^{\dagger}_{m'n'}\,.
\end{flalign}

We now apply the transformation equations just introduced to the system operators 
$A_{qb}$ and $A_{qc}(t)$, where $ A_{qb}= \sigma_{33}-\sigma_{11}$ and $A_{qc}(t)= e^{-i\omega_p t}\sigma^{\vphantom{\dagger}}_{12}+e^{i\omega_p t}\sigma^{\dagger}_{12}$, which appear in the qutrit-bath and qutrit-cavity interaction terms, respectively. Note that, in general, system operators become time dependent in the rotating frame. However, since $A_{qb}$ is diagonal (and therefore commutes with the rotation operator), it remains unchanged under the transformation.
Since $A_{qb}$ and $A_{qc}(t)$ are Hermitian, Eq.~(\ref{eq_appendixB}) can be applied. For $\tilde{A}_{qb}(t)$, we find
\begin{align}
    \tilde{A}_{qb}(t) = \sum_{m,n} e^{i\theta^{\parallel}_{mn}(t)}\xi^{\parallel}_{mn} F_{mn} =  \sum_{m',n'} e^{-i\theta^{\parallel}_{m'n'}(t)}(\xi^{\parallel}_{m'n'})^* F^{\dagger}_{m'n'}\,,
    \label{eq_bsysop}
\end{align}
with 
\begin{align}
\xi^{\parallel}_{mn}
&= \abs{\mel{\mu_m}{\sigma_{33}-\sigma_{11}}{\mu_n}}
   e^{i\varsigma^{\parallel}_0}\,,
\nonumber\\
\xi^{\parallel}
&=
\begin{pmatrix}
-\cos(2\zeta)
&
-\sin(2\zeta)\sin\phi
&
-\sin(2\zeta)\cos\phi
\\[6pt]
-\sin(2\zeta)\sin\phi
&
\cos(2\zeta)\sin^2\phi
&
\dfrac{1}{2}\cos(2\zeta)\sin(2\phi)
\\[6pt]
-\sin(2\zeta)\cos\phi
&
\dfrac{1}{2}\cos(2\zeta)\sin(2\phi)
&
\cos(2\zeta)\cos^2\phi
\end{pmatrix}\,,
\end{align}
and
\begin{align}
\theta^{\parallel}_{mn}(t)
&= \left(\dot{\alpha}_n-\dot{\alpha}_m\right)t+\varsigma^{\parallel}(t)
 = \left(\dot{\alpha}_n-\dot{\alpha}_m\right)t\,,
\nonumber\\
\theta^{\parallel}(t)
&= 
\begin{pmatrix}
0
&
-\dfrac{\Delta}{2}-\dfrac{\Omega_T}{2}
&
-\dfrac{\Delta}{2}+\dfrac{\Omega_T}{2}
\\[6pt]
\dfrac{\Delta}{2}+\dfrac{\Omega_T}{2}
&
0
&
\Omega_T
\\[6pt]
\dfrac{\Delta}{2}-\dfrac{\Omega_T}{2}
&
-\Omega_T
&
0
\end{pmatrix}\,t\,,
\end{align}
where
\begin{eqnarray}
\Omega_T = \sqrt{\Delta^2 + \Omega_s^2 + \Omega_p^2}\,.
\end{eqnarray}
Note that the matrices $\xi^{\parallel}$ and $\theta^{\parallel}(t)$ use the ordering $|\mu_0\rangle$, $|\mu_+\rangle$, and $|\mu_-\rangle$ of the basis states.
To transform the qutrit--cavity interaction term, we use
\begin{align}
\mel{\mu_m}{A_{qc}(t)}{\mu_n}
&= \mel{\mu_m}{\left(e^{-i\omega_p t}\sigma^{\vphantom{\dagger}}_{12}
         + e^{i\omega_p t}\sigma^{\dagger}_{12}\right)}{\mu_n}
\nonumber\\
&= e^{-i\omega_p t}
   \mel{\mu_m}{\sigma^{\vphantom{\dagger}}_{12}}{\mu_n}
   + e^{i\omega_p t}
   \mel{\mu_m}{\sigma^{\dagger}_{12}}{\mu_n}
\nonumber\\
&= e^{-i\omega_p t}
   \abs{\mel{\mu_m}{\sigma^{\vphantom{\dagger}}_{12}}{\mu_n}}
   e^{i\varsigma^{\perp,12}_0}
\nonumber\\
&\quad
   + e^{i\omega_p t}
   \abs{\mel{\mu_m}{\sigma^{\dagger}_{12}}{\mu_n}}
   e^{i\varsigma^{\perp,21}_0}
\nonumber\\
&= e^{-i\omega_p t} \xi^{\perp,12}_{mn}
   + e^{i\omega_p t} \xi^{\perp,21}_{mn}\,.
\end{align}
It follows 
\begin{align}
\tilde{A}_{qc}(t) &= \sum_{m,n} e^{i(\dot{\alpha}_n-\dot{\alpha}_m)t} \left(e^{-i\omega_p t} \xi^{\perp,12}_{mn} + e^{i\omega_p t}\xi^{\perp,21}_{mn}\right)F_{mn}, \nonumber \\
\tilde{A}_{qc}(t) &= \sum_{m,n} \left(e^{i\theta^{\perp,12}_{mn}(t)} \xi^{\perp,12}_{mn} + e^{i\theta^{\perp,21}_{mn}(t)}\xi^{\perp,21}_{mn}\right)F_{mn}
    \nonumber \\
    &\quad \times \sum_{m',n'} \left(e^{-i\theta^{\perp,12}_{m'n'}(t)} (\xi^{\perp,12}_{m'n'})^* + e^{-i\theta^{\perp,21}_{m'n'}(t)}(\xi^{\perp,21}_{m'n'})^*\right)F^{\dagger}_{m'n'},
\label{eq_csysop}
\end{align}
with
\begin{align}
\xi^{\perp,12}_{mn}
&= \abs{\mel{\mu_m}{\sigma_{12}}{\mu_n}}
   e^{i\varsigma^{\perp,12}_0},
\nonumber\\
\xi^{\perp,12}
&=
{\left(\xi^{\perp,21}\right)}^T =
\begin{pmatrix}
0 & \cos\zeta\cos\phi & -\cos\zeta\sin\phi \\
0 & \sin\zeta\sin\phi\cos\phi & -\sin\zeta\sin^2\phi \\
0 & \sin\zeta\cos^2\phi & -\sin\zeta\sin\phi\cos\phi
\end{pmatrix}\,,
\label{eq_xiperp}\\
\theta^{\perp,12}_{mn}(t)
&= \alpha_n(t)-\alpha_m(t)+\varsigma^{\perp,12}(t)
   = \left(\dot{\alpha}_n-\dot{\alpha}_m-\omega_p\right)t,
\nonumber\\
\theta^{\perp,12}(t)
&=
\begin{pmatrix}
-\omega_p
&
-\dfrac{\Delta}{2}-\dfrac{\Omega_T}{2}-\omega_p
&
-\dfrac{\Delta}{2}+\dfrac{\Omega_T}{2}-\omega_p
\\[6pt]
\dfrac{\Delta}{2}+\dfrac{\Omega_T}{2}-\omega_p
&
-\omega_p 
&
\Omega_T-\omega_p
\\[6pt]
\dfrac{\Delta}{2}-\dfrac{\Omega_T}{2}-\omega_p
&
-\Omega_T-\omega_p
&
-\omega_p 
\end{pmatrix}\,t\,,
\nonumber\\[8pt]
\theta^{\perp,21}_{mn}(t)
&= \alpha_n(t)-\alpha_m(t)+\varsigma^{\perp,21}(t)
   = \left(\dot{\alpha}_n-\dot{\alpha}_m+\omega_p\right)t \,,
\nonumber\\
\theta^{\perp,21}(t)
&=
\begin{pmatrix}
\omega_p
&
-\dfrac{\Delta}{2}-\dfrac{\Omega_T}{2}+\omega_p
&
-\dfrac{\Delta}{2}+\dfrac{\Omega_T}{2}+\omega_p
\\[6pt]
\dfrac{\Delta}{2}+\dfrac{\Omega_T}{2}+\omega_p
&
\omega_p
&
\Omega_T+\omega_p
\\[6pt]
\dfrac{\Delta}{2}-\dfrac{\Omega_T}{2}+\omega_p
&
-\Omega_T+\omega_p
&
\omega_p
\end{pmatrix}\,t\,.
\end{align}

Now that we have transformed the relevant operators to the interaction picture, we can write down the qutrit-bath interaction Hamiltonian $\tilde{H}_{qb}(t)$ in the interaction picture, 
\begin{align}
\tilde{H}_{qb}(t) 
&=\sum_k g_{\parallel,k} U^{\dagger}_b(t)U^{\dagger}_c(t)U^{\dagger}_q(t) 
\nonumber\\& \qquad\qquad\qquad \times \left(\sigma_{33}-\sigma_{11}\right)\left(b^{\dagger}_k+b^{\vphantom{\dagger}}_k\right)U_q(t)U_c(t)U_b(t) 
\nonumber\\&= \sum_k g_{\parallel,k} U^{\dagger}_q(t)  \left(\sigma_{33}-\sigma_{11}\right)U_q(t)U^{\dagger}_b(t)\left(b^{\dagger}_k+b^{\vphantom{\dagger}}_k\right)U_b(t)
\nonumber\\&= \sum_k g_{\parallel,k}  \tilde{A}_{qb}(t)\left[\prod_{l,j} e^{i\omega_{b,l}b^{\dagger}_lb^{\vphantom{\dagger}}_lt}\left(b^{\dagger}_k+b^{\vphantom{\dagger}}_k\right)e^{-i\omega_{b,j}b^{\dagger}_jb^{\vphantom{\dagger}}_jt}\right]
\nonumber\\&= \left[\sum_{m,n} e^{i\theta^{\parallel}_{mn}(t)}\xi^{\parallel}_{mn} F_{mn}\right]\left[\sum_k g_{\parallel,k} \left(e^{i\omega_{b,k} t}b^{\dagger}_k+e^{-i\omega_{b,k} t}b^{\vphantom{\dagger}}_k\right)\right]\,. 
    \label{eq_hqbint}
\end{align}
Recognizing that the first term in the square brackets in the final line with its Hermitian conjugate coincides with $\tilde{A}_{qb}(t)$ in Eq.~(\ref{eq_bsysop}) and defining
\begin{align}
    \tilde{B}(t)=\sum_k g_{\parallel,k} \left(e^{i\omega_{b,k} t}b^{\dagger}_k+e^{-i\omega_{b,k} t}b^{\vphantom{\dagger}}_k\right)\,,
    \label{eq_bathop}
\end{align}
we obtain
\begin{align}
    \tilde{H}_{qb}(t) =\tilde{A}_{qb}(t)\otimes \tilde{B}(t)\,.
\end{align}

To treat the cavity--bath interaction, we work in a limiting regime. Specifically, we assume that the cavity--bath coupling strengths are much larger than the cavity-qutrit coupling strength, which---in turn---is much larger than the bath-qutrit coupling strengths ($g_{cb,k}\gg g\gg g_{\parallel,k}$). In this regime, the cavity loses energy to the bath at a much faster timescale than it exchanges energy with the qutrit. As a consequence, the coherent exchange between the cavity and the qutrit can be treated as being effectively incoherent and the cavity can be considered to have reached a stationary steady state on the timescales relevant to the qutrit dynamics. 
Effectively, the strongly damped cavity mode exhibits a broadened spectral response that serves as an effective (second) bath for the qutrit. We account for the mode broadening by treating the cavity degrees of freedom as a collection of modes. Assuming $g_{\parallel,k}\ll g\ll g_{cb,k}$, where $g_{\parallel,k}$, $g$, and $g_{cb,k}$ are defined in Eqs.~(\ref{eq_Hqb}), (\ref{eq_Hqc}), and (\ref{eq_Hcb}), respectively, the cavity Hamiltonian $H_{c,R}$ and the qutrit-cavity interaction Hamiltonian $H_{qc,R}(t)$ in the rotating frame read 
\begin{equation}
\label{eq_cavity2}
    H_{c,R} = \sum_k \omega_{c,k} c^{\dagger}_kc^{\vphantom{\dagger}}_k
\end{equation}
and
\begin{equation}
\label{eq_qutirtcavity2}
    H_{qc,R}(t) = \sum_k g_{\perp,k}\left(e^{-i\omega_p t}\sigma^{\vphantom{\dagger}}_{12}+e^{i\omega_p t}\sigma^{\dagger}_{12}\right)\left(c^{\dagger}_k+c^{\vphantom{\dagger}}_k\right)
\end{equation}
while the cavity evolution operator takes the form
\begin{align}
\label{eq_evolcav2}
    U_c(t) =\prod_k e^{-i\omega_{c,k}c^{\dagger}_kc^{\vphantom{\dagger}}_k t}\,.
\end{align}
\begin{widetext}
\noindent The cavity-qutrit interaction Hamiltonian $\tilde{H}_{qc}(t)$ in the interaction picture then reads ($g \rightarrow g_{\perp,k}$)
\begin{align}
    \tilde{H}_{qc}(t) 
    &=\sum_k g_{\perp,k} U^{\dagger}_b(t)U^{\dagger}_c(t)U^{\dagger}_q(t)  \left(e^{-i\omega_p t}\sigma^{\vphantom{\dagger}}_{12}+e^{i\omega_p t}\sigma^{\dagger}_{12}\right)\left(c^{\dagger}+c^{\vphantom{\dagger}}\right)U_q(t)U_c(t)U_b(t) 
    \nonumber\\&=\sum_k g_{\perp,k} U^{\dagger}_q(t)  \left(e^{-i\omega_p t}\sigma^{\vphantom{\dagger}}_{12}+e^{i\omega_p t}\sigma^{\dagger}_{12}\right)U_q(t)U^{\dagger}_c(t)\left(c^{\dagger}+c^{\vphantom{\dagger}}\right)U_c(t)
    \nonumber\\ &= \sum_k g_{\perp,k} \tilde{A}_{qc}(t)\left[\textstyle\prod\limits_{j,l} e^{i\omega_{c,l}c^{\dagger}_lc^{\vphantom{\dagger}}_lt}\left(c^{\dagger}_k+c^{\vphantom{\dagger}}_k\right)e^{-i\omega_{c,j}c^{\dagger}_jc^{\vphantom{\dagger}}_jt}\right]
    \nonumber\\&=   \left[\sum_{m,n} \left(e^{i\theta^{\perp,12}_{mn}(t)} \xi^{\perp,12}_{mn} + e^{i\theta^{\perp,21}_{mn}(t)}\xi^{\perp,21}_{mn}\right)F_{mn}\right]\left[\sum_k g_{\perp,k}\left(e^{i\omega_{c,k} t}c^{\dagger}_k+e^{-i\omega_{c,k} t}c^{\vphantom{\dagger}}_k\right) \right]\,.
    \label{eq_hqcint}
\end{align}
\end{widetext}
Recognizing that the first term in the final line with its Hermitian conjugate coincides with $\tilde{A}_{qc}(t)$ in Eq.~(\ref{eq_csysop}) and defining
\begin{align}
    \tilde{C}(t) =
   \sum_k g_{\perp,k}\left(e^{i\omega_{c,k} t}c^{\dagger}_k+e^{-i\omega_{c,k} t}c^{\vphantom{\dagger}}_k\right) \,,
\end{align}
we obtain
\begin{align}
    \tilde{H}_{qc}(t) =\tilde{A}_{qc}(t)\otimes \tilde{C}(t)\,.
\end{align}
Since the primary role of the cavity--bath interaction Hamiltonian $H_{cb}$ is to establish a ``stationary dissipative character'' of the cavity, it is accounted for implicitly in what follows. Specifically, $H_{cb}$ is incorporated effectively through the cavity decay rate $\kappa_c$, which sets
 the spectral width $\kappa_{\perp}$ of the cavity modes (transverse coupling),
\begin{eqnarray}
\label{eq_kappaeff}
   \kappa_{\perp} =\kappa_c = {2\pi \sum\limits_k g^2_{cb,k} \delta\left(\omega-\omega_{k}\right)} \biggr\rvert_{\omega = \omega_c}\,.
\end{eqnarray}
Equation~(\ref{eq_kappaeff}) can be obtained by deriving an effective master equation for the cavity, which retains only the cavity--bath interaction~\cite{puri2001}.

In this limit, the cavity dynamics are fast compared to the qutrit, allowing for the cavity to be adiabatically eliminated. Because the cavity rapidly reaches its steady state, the cavity annihilation and creation operators that appear in the system-cavity coupling Hamiltonian can be replaced by their steady-state response, yielding the effective structured reservoir used in our study.

We emphasize that this approach constitutes an effective ``cavity-as-bath reduction'' rather than a fully microscopic dynamical treatment of a coherently coupled qutrit-cavity system. This effective treatment is justified under a strict separation of timescales, i.e., when the cavity relaxation rate is much larger than the atom-cavity coupling strength($\kappa_c \gg g$). In this limit, the cavity dynamics are fast compared to the qutrit, allowing for the cavity to be adiabatically eliminated. Because the cavity rapidly reaches its steady state, the cavity annihilation and creation operators that appear in the system-cavity coupling Hamiltonian can be replaced by their steady-state response, yielding the effective structured reservoir used in this study. The parameter $\kappa_{\perp}$ [see Eq.~(\ref{eq_kappaeff})] will be used in Eq.~(\ref{eq_speccav2}) to define the spectral density of the cavity.
In summary, Eqs.~(\ref{eq_hqbint}) and (\ref{eq_hqcint}) provide us with the Hamiltonian $\tilde{H}(t)$ in the interaction picture,
\begin{align}
    \tilde{H}(t) = \tilde{H}_{qb}(t) + \tilde{H}_{qc}(t)\,.
\end{align}
Note that in the interaction picture the other terms of the Hamiltonian go away by construction. 

Using $\tilde{H}(t)$, the reduced qutrit dynamics are obtained from the von Neumann equation [Eq.~(\ref{eq_vN})] by (a) applying the Born approximation ($g_{\perp,k}$ and $g_{\parallel,k}$ much smaller than $\omega_1$ and $\omega_3$, respectively) and the Markovian approximation ($g_{\perp,k}$ and $g_{\parallel,k}$ are much smaller than $\omega_c$ and $\omega_{b,k}$, respectively); (b) making a product ansatz for the density matrix $\rho(t)$ and assuming that the effective bath density matrix $\rho_c$, which accounts for the cavity and the actual bath density matrix $\rho_b$ are stationary, $\tilde{\rho}(t) = \tilde{\rho}_q(t)\otimes\rho_c\otimes \rho_b$; and (c) tracing out the cavity and bath degrees of freedom.
As a result, we obtain the Redfield master equation:
\begin{widetext}
\begin{align}
    \frac{d}{dt}\tilde{\rho}_q(t) &= - \textstyle\int\limits_0^{\infty} ds\text{Tr}_{b,c}\left\{\comm{\tilde{H}_{qb}(t) + \tilde{H}_{qc}(t)}{\comm{\tilde{H}_{qb}(t-s) + \tilde{H}_{qc}(t-s)}{\tilde{\rho}_q(t)\otimes\rho_c\otimes \rho_b}}\right\}
    \nonumber\\&= - \textstyle\int\limits_0^{\infty} ds\text{Tr}_{b}\left\{\comm{\tilde{H}_{qb}(t) }{\comm{\tilde{H}_{qb}(t-s) }{\tilde{\rho}_q(t)\otimes \rho_b}}\right\} - \textstyle\int\limits_0^{\infty} ds\text{Tr}_{c}\left\{\comm{\tilde{H}_{qc}(t)}{\comm{\tilde{H}_{qc}(t-s)}{\tilde{\rho}_q(t)\otimes\rho_c}}\right\}+{\cal{O}}(\text{``mixed terms''})
\,.
\label{eq_ME_og}
\end{align}
The quantity ${\cal{O}}(\text{``mixed terms''})$ collects mixed terms that contain integrands that are of the form $\comm{\tilde{H}_{qb}(t) }{\comm{\tilde{H}_{qc}(t-s) }{\tilde{\rho}_q(t)\otimes \rho_c \otimes \rho_b}}$ and $\comm{\tilde{H}_{qc}(t) }{\comm{\tilde{H}_{qb}(t-s) }{\tilde{\rho}_q(t)\otimes \rho_c \otimes \rho_b}}$. For thermal reservoirs with random phases, the single-operator expectation values vanish, $\expval{b_k^{\vphantom{\dagger}}}=\expval{b_k^{\dagger}} =0$ and $\expval{c_k^{\vphantom{\dagger}}} = \expval{c_k^{\dagger}}=0$; consequently, the mixed terms vanish~\cite{puri2001}.
Using Eqs.~(\ref{eq_hqbint}) and (\ref{eq_hqcint}), Eq.~(\ref{eq_ME_og}) becomes
\begin{align}
     \frac{d}{dt}\tilde{\rho}_q(t) &= - {\textstyle\int\limits_0^{\infty}} ds\text{Tr}_{b}\left\{\comm{\tilde{H}_{qb}(t) }{\comm{\tilde{H}_{qb}(t-s) }{\tilde{\rho}_q(t)\otimes \rho_b}}\right\} - {\textstyle\int\limits_0^{\infty}} ds\text{Tr}_{c}\left\{\comm{\tilde{H}_{qc}(t)}{\comm{\tilde{H}_{qc}(t-s)}{\tilde{\rho}_q(t)\otimes\rho_c}}\right\}\qquad\qquad\qquad\quad
    \nonumber\\&= {\textstyle\int\limits_0^{\infty}} ds\text{Tr}_{b}\left\{\tilde{H}_{qb}(t) \tilde{\rho}_q(t)\otimes \rho_b\tilde{H}_{qb}(t-s)\right\}- {\textstyle\int\limits_0^{\infty}} ds \text{Tr}_{b}\left\{\tilde{H}_{qb}(t) \tilde{H}_{qb}(t-s)\tilde{\rho}_q(t)\otimes \rho_b \right\}
    \nonumber\\& \qquad+ {\textstyle\int\limits_0^{\infty}} ds\text{Tr}_{b}\left\{\tilde{H}_{qb}(t-s) \tilde{\rho}_q(t)\otimes \rho_b\tilde{H}_{qb}(t)\right\}- {\textstyle\int\limits_0^{\infty}} ds \text{Tr}_{b}\left\{\tilde{\rho}_q(t)\otimes \rho_b\tilde{H}_{qb}(t-s) \tilde{H}_{qb}(t)\right\}
    \nonumber\\&\qquad+ {\textstyle\int\limits_0^{\infty}} ds\text{Tr}_{c}\left\{\tilde{H}_{qc}(t) \tilde{\rho}_q(t)\otimes \rho_c\tilde{H}_{qc}(t-s)\right\}- {\textstyle\int\limits_0^{\infty}} ds \text{Tr}_{c}\left\{\tilde{H}_{qc}(t) \tilde{H}_{qc}(t-s)\tilde{\rho}_q(t)\otimes \rho_c \right\}
    \nonumber\\& \qquad+ {\textstyle\int\limits_0^{\infty}} ds\text{Tr}_{c}\left\{\tilde{H}_{qc}(t-s) \tilde{\rho}_q(t)\otimes \rho_c\tilde{H}_{qc}(t)\right\}- {\textstyle\int\limits_0^{\infty}} ds \text{Tr}_{c}\left\{\tilde{\rho}_q(t)\otimes \rho_c\tilde{H}_{qc}(t-s) \tilde{H}_{qc}(t)\right\}\nonumber
\end{align}
\begin{align}
    &= {\textstyle\int\limits_0^{\infty}} ds \tilde{A}_{qb}(t) \tilde{\rho}_q(t) \tilde{A}_{qb}(t-s) \text{Tr}_{b}\left\{ \tilde{B}(t) \rho_b \tilde{B}(t-s)\right\}- {\textstyle\int\limits_0^{\infty}} ds \tilde{A}_{qb}(t) \tilde{A}_{qb}(t-s) \tilde{\rho}_q(t)  \text{Tr}_{b}\left\{ \tilde{B}(t) \tilde{B}(t-s)\rho_b \right\}
    \nonumber\\& \qquad+ {\textstyle\int\limits_0^{\infty}} ds \tilde{A}_{qb}(t-s) \tilde{\rho}_q(t) \tilde{A}_{qb}(t) \text{Tr}_{b}\left\{ \tilde{B}(t-s) \rho_b \tilde{B}(t)\right\}- {\textstyle\int\limits_0^{\infty}} ds \tilde{\rho}_q(t) \tilde{A}_{qb}(t-s) \tilde{A}_{qb}(t)   \text{Tr}_{b}\left\{ \rho_b\tilde{B}(t-s) \tilde{B}(t) \right\}
    \nonumber\\&\qquad+ {\textstyle\int\limits_0^{\infty}} ds \tilde{A}_{qc}(t) \tilde{\rho}_q(t) \tilde{A}_{qc}(t-s) \text{Tr}_{c}\left\{ \tilde{C}(t) \rho_c \tilde{C}(t-s)\right\}- {\textstyle\int\limits_0^{\infty}} ds \tilde{A}_{qc}(t) \tilde{A}_{qc}(t-s) \tilde{\rho}_q(t)  \text{Tr}_{c}\left\{ \tilde{C}(t) \tilde{C}(t-s)\rho_c \right\}
    \nonumber\\& \qquad + {\textstyle\int\limits_0^{\infty}} ds \tilde{A}_{qc}(t-s) \tilde{\rho}_q(t) \tilde{A}_{qc}(t) \text{Tr}_{c}\left\{ \tilde{C}(t-s) \rho_c \tilde{C}(t)\right\}- {\textstyle\int\limits_0^{\infty}} ds \tilde{\rho}_q(t) \tilde{A}_{qc}(t-s) \tilde{A}_{qc}(t)   \text{Tr}_{c}\left\{ \rho_c\tilde{C}(t-s) \tilde{C}(t) \right\}\,.
\end{align}
Using the cyclic property of the trace and the fact that the bath and cavity bath density matrices $\rho_b$ and $\rho_c$ are time independent, it can be shown that the two-time bath and cavity-bath correlation functions have the following properties: 
\begin{align}
\label{eq_corrbath}
    &\text{Tr}_b\left\{\tilde{B}(t)\tilde{B}(t-s) \rho_b\right\} = \expval{\tilde{B}(t)\tilde{B}(t-s)} = \expval{\tilde{B}(s)\tilde{B}(0)}, \quad \text{Tr}_b\left\{\tilde{B}(t-s)\tilde{B}(t) \rho_b\right\} = \expval{\tilde{B}(0)\tilde{B}(s)}\,,
    \\&
    \label{eq_corrbathcavity}\text{Tr}_c\left\{\tilde{C}(t)\tilde{C}(t-s) \rho_c\right\} =\expval{\tilde{C}(s)\tilde{C}(0)}, \quad \text{Tr}_c\left\{\tilde{C}(t-s)\tilde{C}(t) \rho_c\right\} =\expval{\tilde{C}(0)\tilde{C}(s)}\,.
\end{align}
Using Eqs.~(\ref{eq_corrbath}) and (\ref{eq_corrbathcavity}) and rearranging terms, the master equation simplifies to
\begin{align}
\frac{d}{dt}\tilde{\rho}_q(t) = &{\int\limits_0^{\infty}} ds \left[\left(\tilde{A}_{qb}(t) \tilde{\rho}_q(t) \tilde{A}_{qb}(t-s) - \tilde{\rho}_q(t) \tilde{A}_{qb}(t-s) \tilde{A}_{qb}(t)\right)   \expval{\tilde{B}(0)\tilde{B}(s)}+ \left(\tilde{A}_{qb}(t-s) \tilde{\rho}_q(t) \tilde{A}_{qb}(t) - \tilde{A}_{qb}(t) \tilde{A}_{qb}(t-s) \tilde{\rho}_q(t)\right)  \expval{\tilde{B}(s)\tilde{B}(0)}\right]
\nonumber\\+ &{\int\limits_0^{\infty}} ds \left[\left(\tilde{A}_{qc}(t) \tilde{\rho}_q(t) \tilde{A}_{qc}(t-s) - \tilde{\rho}_q(t) \tilde{A}_{qc}(t-s) \tilde{A}_{qc}(t) \right)  \expval{\tilde{C}(0)\tilde{C}(s)}
+ \left(\tilde{A}_{qc}(t-s) \tilde{\rho}_q(t) \tilde{A}_{qc}(t) - \tilde{A}_{qc}(t) \tilde{A}_{qc}(t-s) \tilde{\rho}_q(t) \right)\expval{\tilde{C}(s)\tilde{C}(0)}\right]
\nonumber\\= &{\int\limits_0^{\infty}} ds \left[\left(\tilde{A}_{qb}(t) \tilde{\rho}_q(t) \tilde{A}_{qb}(t-s) - \tilde{\rho}_q(t) \tilde{A}_{qb}(t-s) \tilde{A}_{qb}(t)\right)   [\Lambda_b(s)]^*+ \left(\tilde{A}_{qb}(t-s) \tilde{\rho}_q(t) \tilde{A}_{qb}(t) - \tilde{A}_{qb}(t) \tilde{A}_{qb}(t-s) \tilde{\rho}_q(t)\right) \Lambda_b(s)\right]
\nonumber\\+ &{\int\limits_0^{\infty}} ds \left[\left(\tilde{A}_{qc}(t) \tilde{\rho}_q(t) \tilde{A}_{qc}(t-s) - \tilde{\rho}_q(t) \tilde{A}_{qc}(t-s) \tilde{A}_{qc}(t) \right)  [\Lambda_c(s)]^*
+ \left(\tilde{A}_{qc}(t-s) \tilde{\rho}_q(t) \tilde{A}_{qc}(t) - \tilde{A}_{qc}(t) \tilde{A}_{qc}(t-s) \tilde{\rho}_q(t) \right)\Lambda_c(s)\right]\,.
\label{eq_MEx}
\end{align}
\end{widetext}
In the last equality of Eq.~(\ref{eq_MEx}), we introduced the bath correlation 
function $\Lambda_b(s)$,
\begin{align}
\Lambda_b(s) = \expval{\tilde{B}(s)\tilde{B}(0)}\,.
\end{align}
Inserting
the definition of $\tilde{B}(s)$, 
Eq.~(\ref{eq_ham_bath}), we find
\begin{align}
       \Lambda_b(s) &=\sum\limits_{k,k'} g_{\parallel,k}g_{\parallel,k'} \expval{\left(e^{i\omega_{b,k} s}b^{\dagger}_{k}+e^{-i\omega_{b,k}s}b^{\vphantom{\dagger}}_{k}\right)\left(b^{\dagger}_{k'}+b^{\vphantom{\dagger}}_{k'}\right)}\nonumber\\
        &=\sum\limits_{k,k'} g_{\parallel,k}g_{\parallel,k'} \bigg(
        e^{i\omega_{b,k} s}\expval{b^{\dagger}_kb^{\dagger}_{k'}} + e^{-i\omega_{b,k} s}\expval{b_kb^{\dagger}_{k'}} 
        \nonumber \\&\qquad\qquad\qquad+ e^{i\omega_{b,k} s}\expval{b^{\dagger}_kb_{k'}} + e^{i\omega_{b,k} s}\expval{b_kb_{k'}}\bigg)\,.
    \label{eq_lab_b}
\end{align}
The correlators 
$\expval{b^{\vphantom{\dagger}}_kb^{\vphantom{\dagger}}_{k'}}$ and $ \expval{b^{\dagger}_kb^{\dagger}_{k'}}$ in Eq.~(\ref{eq_lab_b}) are,
consistent with a thermal bath with random phases~\cite{puri2001}, taken to be zero. Under this assumption, we have 
\begin{align}
\begin{split}
\Lambda_b(s) &=\sum\limits_{k,k'} g_{\parallel,k}g_{\parallel,k'} \left(
e^{-i\omega_{b,k} s}\expval{b_kb^{\dagger}_{k'}} + e^{i\omega_{b,k} s}\expval{b^{\dagger}_kb_{k'}}\right)
\\ &=\sum\limits_{k,k'} g_{\parallel,k}g_{\parallel,k'} \left[
e^{-i\omega_{b,k} s}\left(1+\expval{n_{b,k}}\right)\delta_{k,k'} + e^{i\omega_{b,k} s}\expval{n_{b,k}}\delta_{k,k'} \right]
\\ &= \sum\limits_{k} g^2_{\parallel,k} \left[
e^{-i\omega_{b,k'} s}\left(1+\expval{n_{b,k}}\right) + e^{i\omega_{b,k'} s}\expval{n_{b,k}} \right]\,.
\end{split}
\end{align}
The expectation value $\expval{n_{b,k}}$, i.e., the occupation number of the $k$th bath mode, 
 is given by the Planck distribution $\expval{n_{b,k}}$,
\begin{align}
    \expval{n_{b,k}}={\left[e^{ \omega_{b,k}/(k_{\text{B}}T_b)}-1\right]}^{-1}\,,
\end{align}
where $T_b$ and $k_B$ denote the bath temperature and Boltzmann constant, respectively. To switch from discrete modes to a continuum of modes, it is useful to define $n_b(\omega)$,
\begin{eqnarray}
n_b(\omega)= {\left[e^{ \omega/(k_{\text{B}}T_b)}-1\right]}^{-1}\,.
\end{eqnarray}  
With this, we have
\begin{align}
    \Lambda_{b}(s) = {\int\limits_0^{\infty}} d\omega J_{\parallel}(\omega) \left\{e^{-i\omega s}\left[1+n_b(\omega)\right]+e^{i\omega s}n_b(\omega)  \right\}\,,
    \label{eq_lambath}
\end{align}
where the spectral density function $J_{\parallel}(\omega)$ for the longitudinal coupling is
\begin{align}
    J_{\parallel}(\omega) = \sum_k g^2_{\parallel,k} \delta\left(\omega-\omega_{b,k}\right)= \dfrac{\Gamma_{\parallel}}{\pi}\dfrac{(\kappa_{\parallel}/2)^2}{(\kappa_{\parallel}/2)^2 +\omega^2}\,\vcenter{\hbox{.}}
    \label{eq_specbath}
\end{align}
Here, $\kappa_{\parallel}$ denotes the spectral width of the bath, which corresponds to the full width at half maximum of the coupling strength distribution, as encoded in the spectral density. The quantity $\Gamma_{\parallel}$ is equal to the maximum dephasing rate or, equivalently, the rate on resonance, where the system frequency matches the bath resonance frequency,
\begin{eqnarray}
    \Gamma_{\parallel} = \pi\left[\sum_k g^2_{\parallel,k} \delta\left(\omega-\omega_{b,k}\right) \Bigg|_{\omega = 0}\right]\,.
\end{eqnarray}
Similarly, we have for 
the cavity correlation function $\Lambda_{c}(s)$ (transverse coupling), 
\begin{align}
    \text{}\;\; \Lambda_{c}(s)= {\int\limits_0^{\infty}} d\omega J_{\perp}(\omega) \left\{e^{-i\omega s}\left[1+n_c(\omega)\right]+e^{i\omega s}n_c(\omega)  \right\}\,,
    \label{eq_lamcav}
\end{align}
where
\begin{align}
   n_{c,k}=\dfrac{1}{e^{\omega_{c,k}/(k_{\text{B}}T_c)}-1}
    \label{eq_speccav1}
\end{align} 
and
\begin{align}
    J_{\perp}(\omega) = \dfrac{\Gamma_{\perp}}{\pi}\dfrac{\left(\kappa_{\perp}/2\right)^2}{\left(\kappa_{\perp}/2\right)^2 + \left(\omega-\omega_c\right)^2}\,\vcenter{\hbox{,}}
    \label{eq_speccav2}
\end{align} 
where $\Gamma_{\perp} = 2g^2/\kappa_{\perp}$. Since bath and cavity operators in the interaction Hamiltonians are Hermitian, we have $\langle\tilde{B}(0)\tilde{B}(s)\rangle=\langle\tilde{B}(s)\tilde{B}(0)\rangle^* = [\Lambda_b(s)]^* \text{~and~} \langle\tilde{C}(0)\tilde{C}(s)\rangle=\langle\tilde{C}(s)\tilde{C}(0)\rangle^* = [\Lambda_c(s)]^*$.
Inserting $\tilde{A}_{qb}(t)$, $\tilde{A}_{qc}(t)$, $\tilde{A}_{qb}(t-s)$, and $\tilde{A}_{qc}(t-s)$ and their Hermitian conjugates into Eq.~(\ref{eq_MEx}), we have
\begin{widetext}
\begin{align}
    \frac{d}{dt}\tilde{\rho}_q(t) = &\textstyle\int\limits_0^{\infty} ds \left[\left(\sum\limits_{m,n} e^{i\theta^{\parallel}_{mn}(t)}\xi^{\parallel}_{mn} F_{mn}\right) \tilde{\rho}_q(t) \left(\sum\limits_{m',n'} e^{-i\theta^{\parallel}_{m'n'}(t-s)}\xi^{\parallel}_{m'n'} F^{\dagger}_{m'n'}\right)- \tilde{\rho}_q(t)\left(\sum\limits_{m',n'} e^{-i\theta^{\parallel}_{m'n'}(t-s)}\xi^{\parallel}_{m'n'} F^{\dagger}_{m'n'}\right) \left(\sum\limits_{m,n} e^{i\theta^{\parallel}_{mn}(t)}\xi^{\parallel}_{mn} F_{mn}\right)\right]   [\Lambda_b(s)]^*
    \nonumber\\
    + &\textstyle\int\limits_0^{\infty} ds \left[ \left(\sum\limits_{m',n'} e^{i\theta^{\parallel}_{m'n'}(t-s)}\xi^{\parallel}_{m'n'} F_{m'n'}\right) \tilde{\rho}_q(t) \left(\sum\limits_{m,n} e^{-i\theta^{\parallel}_{mn}(t)}\xi^{\parallel}_{mn} F^{\dagger}_{mn}\right)- \left(\sum\limits_{m,n} e^{-i\theta^{\parallel}_{mn}(t)}\xi^{\parallel}_{mn} F^{\dagger}_{mn}\right)  \left(\sum\limits_{m',n'} e^{i\theta^{\parallel}_{m'n'}(t-s)}\xi^{\parallel}_{m'n'} F_{m'n'}\right) \tilde{\rho}_q(t)\right] \Lambda_b(s)
    \nonumber\\
    + &\textstyle\int\limits_0^{\infty} ds \left[\left(\sum_{\substack{\upsilon=\{12,21\},  m,n}} e^{i\theta^{\perp,\upsilon}_{mn}(t)} \xi^{\perp,\upsilon}_{mn}F_{mn}\right)\tilde{\rho}_q(t)  \left(\sum_{\substack{\upsilon'=\{12,21\},  m',n'}} e^{-i\theta^{\perp,\upsilon'}_{m'n'}(t-s)} (\xi^{\perp,\upsilon'}_{m'n'})^* F^{\dagger}_{m'n'}\right) \right] [\Lambda_c(s)]^*
    \nonumber\\
    - &\textstyle\int\limits_0^{\infty} ds \left[\tilde{\rho}_q(t) \left(\sum_{\substack{\upsilon'=\{12,21\},  m',n'}} e^{-i\theta^{\perp,\upsilon'}_{m'n'}(t-s)} (\xi^{\perp,\upsilon'}_{m'n'})^* F^{\dagger}_{m'n'}\right)\left(\sum_{\substack{\upsilon=\{12,21\},  m,n}} e^{i\theta^{\perp,\upsilon}_{mn}(t)} \xi^{\perp,\upsilon}_{mn}F_{mn}\right) \right]  [\Lambda_c(s)]^*
\nonumber\\
+ &\textstyle\int\limits_0^{\infty} ds \left[ \left(\textstyle\sum_{\substack{\upsilon'=\{12,21\}, m',n'}} e^{i\theta^{\perp,\upsilon'}_{m'n'}(t-s)} \xi^{\perp,\upsilon'}_{m'n'} F_{m'n'}\right) \tilde{\rho}_q(t) \left(\textstyle\sum_{\substack{\upsilon=\{12,21\}, m,n}} e^{-i\theta^{\perp,\upsilon}_{mn}(t)} \xi^{\perp,\upsilon}_{mn}F^{\dagger}_{mn}\right) \right] \Lambda_c(s)
\nonumber\\
- &\textstyle\int\limits_0^{\infty} ds \left[\left(\textstyle\sum_{\substack{\upsilon=\{12,21\},  m,n}} e^{-i\theta^{\perp,\upsilon}_{mn}(t)} \xi^{\perp,\upsilon}_{mn}F^{\dagger}_{mn}\right) \left(\textstyle\sum_{\substack{\upsilon'=\{12,21\},  m',n'}} e^{i\theta^{\perp,\upsilon'}_{m'n'}(t-s)} \xi^{\perp,\upsilon'}_{m'n'} F_{m'n'}\right)\tilde{\rho}_q(t) \right]\Lambda_c(s)\,.
\label{eq_mastereqref}
\end{align}
To simplify, we define the following instantaneous frequencies: 
\begin{align}
    &\alpha^{\beta}_{mn} = -\dfrac{d\theta^{\beta}_{mn}(t)}{dt}\,, \quad \alpha^{\parallel}_{mn} = \dot{\alpha}_m-\dot{\alpha}_n\,, \quad \alpha^{\perp,12}_{mn} = \dot{\alpha}_m-\dot{\alpha}_n + \omega_p\,, \quad \alpha^{\perp,21}_{mn} = \dot{\alpha}_m-\dot{\alpha}_n - \omega_p\,.
    \label{eq_IF}
\end{align}
As the phases $\theta^{\beta}_{mn}(t)$ depend linearly on time, the instantaneous frequencies are constant. Substituting Eqs.~(\ref{eq_thetaphase}) and (\ref{eq_IF}) into the master equation [Eq.~(\ref{eq_mastereqref})], we arrive at
\begin{align}
    \frac{d}{dt}\tilde{\rho}_q(t) = &\sum_{m,n,m',n'} \left\{\exp[-i\left(\alpha^{\parallel}_{mn}-\alpha^{\parallel}_{m'n'}\right)t]\xi^{\parallel}_{mn}\xi^{\parallel}_{m'n'} \int_0^{\infty} ds e^{-i\alpha^{\parallel}_{m'n'}s}[\Lambda_b(s)]^*\right\} \left[F_{mn}\tilde{\rho}_q(t)F^{\dagger}_{m'n'} -\tilde{\rho}_q(t)F^{\dagger}_{m'n'}F_{mn}\right]
    \nonumber\\+&\sum_{m,n,m',n'} \left\{\exp[i\left(\alpha^{\parallel}_{mn}-\alpha^{\parallel}_{m'n'}\right)t]\xi^{\parallel}_{mn}\xi^{\parallel}_{m'n'}\int_0^{\infty} ds e^{i\alpha^{\parallel}_{m'n'}s}\Lambda_b(s)\right\}\left[F_{m'n'}\tilde{\rho}_q(t)F^{\dagger}_{mn} -F^{\dagger}_{mn}F_{m'n'}\tilde{\rho}_q(t)\right]    
    \nonumber\\+ &\sum_{m,n,m',n'} \left\{\quad\;\smashoperator{\sum_{\substack{\upsilon\,\in\{12,21\} \\ \upsilon'\in\{12,21\}}}} \exp[-i\left(\alpha^{\perp,\upsilon}_{mn}-\alpha^{\perp,\upsilon'}_{m'n'}\right)t]\xi^{\perp,\upsilon}_{mn}\xi^{\perp,\upsilon'}_{m'n'} \int_0^{\infty} ds e^{-i\alpha^{\perp,\upsilon'}_{m'n'}s}[\Lambda_c(s)]^*\right\} \left[F_{mn}\tilde{\rho}_q(t)F^{\dagger}_{m'n'} -\tilde{\rho}_q(t)F^{\dagger}_{m'n'}F_{mn}\right]
    \nonumber
\end{align} 
\begin{align}
+&\sum_{m,n,m',n'} \left\{\quad\;\smashoperator{\sum_{\substack{\upsilon\,\in\{12,21\} \\ \upsilon'\in\{12,21\}}}} \exp[i\left(\alpha^{\perp,\upsilon}_{mn}-\alpha^{\perp,\upsilon'}_{m'n'}\right)t]\xi^{\perp,\upsilon}_{mn}\xi^{\perp,\upsilon'}_{m'n'}\int_0^{\infty} ds e^{i\alpha^{\perp,\upsilon'}_{m'n'}s}\Lambda_c(s)\right\}\left[F_{m'n'}\tilde{\rho}_q(t)F^{\dagger}_{mn} -F^{\dagger}_{mn}F_{m'n'}\tilde{\rho}_q(t)\right]
    \nonumber\\
    = &\sum_{m,n,m',n'} \tilde{\Gamma}^*_{\parallel,mn,m'n'}(t)\left[F_{mn}\tilde{\rho}_q(t)F^{\dagger}_{m'n'} -\tilde{\rho}_q(t)F^{\dagger}_{m'n'}F_{mn}\right] + \tilde{\Gamma}_{\parallel,mn,m'n'}(t)\left[F_{m'n'}\tilde{\rho}_q(t)F^{\dagger}_{mn} -F^{\dagger}_{mn}F_{m'n'}\tilde{\rho}_q(t)\right]    
    \nonumber\\+ &\sum_{m,n,m',n'} \tilde{\Gamma}^*_{\perp,mn,m'n'}(t)\left[F_{mn}\tilde{\rho}_q(t)F^{\dagger}_{m'n'} -\tilde{\rho}_q(t)F^{\dagger}_{m'n'}F_{mn}\right] + \tilde{\Gamma}_{\perp,mn,m'n'}(t)\left[F_{m'n'}\tilde{\rho}_q(t)F^{\dagger}_{mn} -F^{\dagger}_{mn}F_{m'n'}\tilde{\rho}_q(t)\right]    
    \nonumber\\ = &\sum_{m,n,m',n'} \left(\tilde{\Gamma}_{\parallel,mn,m'n'}(t)+\tilde{\Gamma}_{\perp,mn,m'n'}(t)\right)\left[F_{m'n'}\tilde{\rho}_q(t)F^{\dagger}_{mn} -F^{\dagger}_{mn}F_{m'n'}\tilde{\rho}_q(t)\right]  + \text{H.c.} \,,
\end{align}
where 
\begin{align}
    &\tilde{\Gamma}_{\parallel,mn,m'n'}(t) = \exp[i\left(\alpha^{\parallel}_{mn}-\alpha^{\parallel}_{m'n'}\right)t]\xi^{\parallel}_{mn}\xi^{\parallel}_{m'n'}\int_0^{\infty} ds e^{i\alpha^{\parallel}_{m'n'}s}\Lambda_b(s)\,,
    \\& \tilde{\Gamma}_{\perp,mn,m'n'}(t) = 
    \smashoperator{\sum_{\substack{\upsilon\,\in\{12,21\} \\ \upsilon'\in\{12,21\}}}}
    \exp[i\left(\alpha^{\perp,\upsilon}_{mn}-\alpha^{\perp,\upsilon'}_{m'n'}\right)t]\xi^{\perp,\upsilon}_{mn}\xi^{\perp,\upsilon'}_{m'n'}\int_0^{\infty} ds e^{i\alpha^{\perp,\upsilon'}_{m'n'}s}\Lambda_c(s)\,.
\end{align}

With the master equation in the interaction picture obtained in its final form, we transform back to the Schr\"odinger picture.
The master equation in the Schr\"odinger picture reads 
\begin{align}
       \dfrac{d}{dt}\rho_{q,R}(t) &= -i\comm{H_{q,R}(t)}{\rho_{q,R}(t)} + \sum_{m,n,m',n'} \left(\tilde{\Gamma}_{\parallel,mn,m'n'}(t)+\tilde{\Gamma}_{\perp,mn,m'n'}(t)\right)\left[F^s_{m'n'}(t)\rho_{q,R}(t)(F^s_{mn}(t))^{\dagger} -(F^s_{mn}(t))^{\dagger}F^s_{m'n'}(t)\rho_{q,R}(t)\right]  + \text{H.c.}\,,
\end{align}
where the jump operators $F_{mn}^s(t)$ in the Schr\"odinger picture are given by 
\begin{align}
    F^s_{mn}(t) = U(t) F_{mn} U^{\dagger}(t) = \sum_{m',n'} e^{i\dot{\alpha}_{m'} t}\ket{\mu_{m'}}\bra{\mu_{m'}}\ket{\mu_m}\bra{\mu_n}\ket{\mu_{n'} }\bra{\mu_{n'} }e^{-i\dot{\alpha}_{n'} t} = e^{i(\dot{\alpha}_m-\dot{\alpha}_n)t}F_{mn}\,.
\end{align}
Rather than working with explicitly time-dependent jump operators, we find it convenient to absorb the time-dependent phase factors contained in $F_{mn}^s(t)$ into the dissipation coefficients. To this end, we define
\begin{align}
\begin{split}
    & \Gamma_{mn,m'n'}(t) =  \Gamma_{\parallel,mn,m'n'} + \Gamma_{\perp,mn,m'n'}(t)\,,
    \\&\Gamma_{\parallel,mn,m'n'} = e^{i(\dot{\alpha}_m-\dot{\alpha}_n)t} e^{-i(\dot{\alpha}_{m'}-\dot{\alpha}_{n'})t}\tilde{\Gamma}_{\parallel,mn,m'n'}(t) = \xi^{\parallel}_{mn}\xi^{\parallel}_{m'n'}\int_0^{\infty} ds e^{i\alpha^{\parallel}_{m'n'}s}\Lambda_b(s)\,,
    \\& \Gamma_{\perp,mn,m'n'}(t) = e^{i(\dot{\alpha}_m-\dot{\alpha}_n)t} e^{-i(\dot{\alpha}_{m'}-\dot{\alpha}_{n'})t}   \tilde{\Gamma}_{\perp,mn,m'n'}(t) = 
    \smashoperator{\sum_{\substack{\upsilon\,\in\{12,21\} \\ \upsilon'\in\{12,21\}}}}
    e^{i\left(\omega^{\perp,\upsilon}_{d}-\omega^{\perp,\upsilon'}_{d}\right)\,t}\xi^{\perp,\upsilon}_{mn}\xi^{\perp,\upsilon'}_{m'n'}\int_0^{\infty} ds e^{i\alpha^{\perp,\upsilon'}_{m'n'}s}\Lambda_c(s)\,,
\end{split}
\label{eq_disscoeff}
\end{align}
where $\omega^{\perp,12}_d = \omega_p$ and $\omega^{\perp,21}_d = -\omega_p$. Note that the ``final'' longitudinal coupling dissipation coefficients $\Gamma_{\parallel,mn,m'n'}$ are time independent while the transverse coupling dissipation coefficients $\Gamma_{\perp,mn,m'n'}(t)$ are time dependent. The master equation in the rotating frame in terms of the ``final'' dissipation coefficients $\Gamma_{mn,m'n'}(t)$ [the time dependence enters via $\Gamma_{\perp,mn,m'n'}(t)$] is given in Eq.~(\ref{eq_master3}). 
While the derivation of Eq.~(\ref{eq_master3}) is for Application 2, it can be adjusted to apply to Application 1 by setting the coherent coupling $\Omega_s$ to zero and by interpreting the coupling to the cavity as a transverse coupling to the bath ($c_k \rightarrow b_k$). 

We emphasize that our derivation does not make any secular approximations with regards to the driving field amplitudes $\Omega_p$ and $\Omega_s$, i.e., the IME, Eq.~(\ref{eq_master3}), is non-perturbative in $\Omega_p$ and $\Omega_s$. Our derivation does, however, rely on the RWA, which requires $\omega_p \gg |\Omega_p|$ and $\omega_s \gg |\Omega_s|$. This approximation, which is justified for the system under consideration as the driving frequencies ($\sim 100 $ THz) tend to be several orders of magnitude larger than the amplitudes ($\sim $ GHz)~\cite{muller2007,ulrich2011}, can be dropped without appreciably complicating the final expressions. 

\section{\label{sec:appendC}Decay rates and Lamb shifts}
This appendix evaluates the dissipation coefficients by explicitly computing the integrals over 
two-point correlation functions of the bath, yielding decay rates and Lamb shifts. We present full details for the cavity (transverse coupling) and summarize the results for the bath (longitudinal coupling). Using Eqs.~(\ref{eq_lamcav}), (\ref{eq_speccav1}), and (\ref{eq_speccav2}) in Eq.~(\ref{eq_disscoeff}), we have
\begin{align}
    \Gamma_{\perp,mn,m'n'}(t) &= \smashoperator{\sum_{\substack{\upsilon\,\in\{12,21\} \\ \upsilon'\in\{12,21\}}}} e^{i\left(\omega^{\perp,\upsilon}_{d}-\omega^{\perp,\upsilon'}_{d}\right)t}\xi^{\perp,\upsilon}_{mn}\xi^{\perp,\upsilon'}_{m'n'}\int\limits_0^{\infty} ds e^{i\alpha^{\perp,\upsilon'}_{m'n'}s}\Lambda_c(s)
    \nonumber\\&= \smashoperator{\sum_{\substack{\upsilon\,\in\{12,21\} \\ \upsilon'\in\{12,21\}}}} e^{i\left(\omega^{\perp,\upsilon}_{d}-\omega^{\perp,\upsilon'}_{d}\right)t}\xi^{\perp,\upsilon}_{mn}\xi^{\perp,\upsilon'}_{m'n'}\left\{\int\limits_0^{\infty} d\omega J_{\perp}(\omega) \int\limits_0^{\infty} ds \left[e^{i\left(\alpha^{\perp,\upsilon'}_{m'n'}+\omega\right) s}n_{c}(\omega) + e^{i\left(\alpha^{\perp,\upsilon'}_{m'n'}-\omega\right) s}\left(1+n_{c}(\omega)\right)\right]\right\}\,.
\end{align}
For a cavity-bath with temperature $T_c=0$, the quantity $n_c(\omega)$ is zero, yielding
\begin{align}
    \Gamma_{\perp,mn,m'n'}(t) &= \smashoperator{\sum_{\substack{\upsilon\,\in\{12,21\} \\ \upsilon'\in\{12,21\}}}} e^{i\left(\omega^{\perp,\upsilon}_{d}-\omega^{\perp,\upsilon'}_{d}\right)t}\xi^{\perp,\upsilon}_{mn}\xi^{\perp,\upsilon'}_{m'n'}\left[\int\limits_0^{\infty} d\omega J_{\perp}(\omega) \int\limits_0^{\infty} ds  e^{i\left(\alpha^{\perp,\upsilon'}_{m'n'}-\omega\right) s}\right] \qquad\qquad\qquad\qquad\qquad\qquad\quad\;\;
    \nonumber\\&= \smashoperator{\sum_{\substack{\upsilon\,\in\{12,21\} \\ \upsilon'\in\{12,21\}}}}e^{i\left(\omega^{\perp,\upsilon}_{d}-\omega^{\perp,\upsilon'}_{d}\right)t}\xi^{\perp,\upsilon}_{mn}\xi^{\perp,\upsilon'}_{m'n'}\left[R\left(\alpha^{\perp,\upsilon'}_{m'n'}\right) + iI\left(\alpha^{\perp,\upsilon'}_{m'n'}\right)\right]\,,
\label{eq_disscoffcav}
\end{align}
where $R$ and $I$ are functions of $\alpha^{\perp,\upsilon'}_{m'n'}$: $R\left(\alpha^{\perp,\upsilon'}_{m'n'}\right) = \Re\left[\int\limits_0^{\infty} d\omega J_{\perp}(\omega) \int\limits_0^{\infty} ds  e^{i\left(\alpha^{\perp,\upsilon'}_{m'n'}-\omega\right) s}\right]$ and $I\left(\alpha^{\perp,\upsilon'}_{m'n'}\right) = \Im\left[\int\limits_0^{\infty} d\omega J_{\perp}(\omega) \int\limits_0^{\infty} ds  e^{i\left(\alpha^{\perp,\upsilon'}_{m'n'}-\omega\right) s}\right]\,$.
Let $u(s)$ be the Heaviside step function and $\mathcal{F}\{u(s)\}(\omega)$ be its Fourier transform. We then have \cite{bateman1954}
\begin{eqnarray}
\int\limits_0^\infty e^{i \omega s}ds 
 = \int\limits_{-\infty}^\infty u(s)e^{i\omega s}ds = \mathcal{F}\{u(s)\}(\omega)
= \pi \delta(\omega) + \mathcal{P}\left(\dfrac{i}{\omega}\right)\,,
\end{eqnarray}
where ${\cal{P}}(x)$ denotes the principal value of $x$. Using this, we obtain 
\begin{align}R\left(\alpha^{\perp,\upsilon'}_{m'n'}\right) + iI\left(\alpha^{\perp,\upsilon'}_{m'n'}\right) & = \int\limits_0^{\infty} d\omega J_{\perp}(\omega) \int\limits_0^{\infty} ds  e^{i\left(\alpha^{\perp,\upsilon'}_{m'n'}-\omega\right) s}
    \nonumber\\& = \int\limits_0^{\infty} d\omega J_{\perp}(\omega)\left[\pi \delta\left(\alpha^{\perp,\upsilon'}_{m'n'}-\omega\right) +i \mathcal{P}\left(\dfrac{1}{ \alpha^{\perp,\upsilon'}_{m'n'}-\omega}\right)\right]
    \nonumber\\& =\begin{cases}
    \pi J_{\perp}\left(\alpha^{\perp,\upsilon'}_{m'n'}\right) -i \mathcal{P} \left( \int\limits_0^{\infty} d\omega \dfrac{J_{\perp}(\omega)}{\omega - \alpha^{\perp,\upsilon'}_{m'n'}}\right)\,\vcenter{\hbox{,}} &\alpha^{\perp,\upsilon'}_{m'n'}\geq 0  
    \\-i \mathcal{P} \left( \int\limits_0^{\infty} d\omega \dfrac{J_{\perp}(\omega)}{\omega - \alpha^{\perp,\upsilon'}_{m'n'}}\right)\,\vcenter{\hbox{,}} &\alpha^{\perp,\upsilon'}_{m'n'}<0
    \end{cases}\,.
\end{align}
The real, physically significant contribution of this integral is
\begin{align}
    R\left(\alpha^{\perp,\upsilon'}_{m'n'}\right) =
    \begin{cases}
    \pi J_{\perp}\left(\alpha^{\perp,\upsilon'}_{m'n'}\right)\,, &\alpha^{\perp,\upsilon'}_{m'n'}\geq 0  
    \\
    0\,, &\alpha^{\perp,\upsilon'}_{m'n'}<0  
    \end{cases}
\end{align}
while the imaginary part reads
\begin{align}
    I\left(\alpha^{\perp,\upsilon'}_{m'n'}\right) =-\dfrac{1}{\pi}\left(\dfrac{2g^2}{\kappa_{\perp}}\right) ( \kappa_{\perp}/2)^2\mathcal{P} \left[ \int\limits_0^{\infty} d\omega \left(\dfrac{1}{\omega-\alpha^{\perp,\upsilon'}_{m'n'}}\right)\left(\dfrac{1}{( \kappa_{\perp}/2)^2+(\omega-\omega_{c})^2}\right) \right] \,\vcenter{\hbox{.}}
\end{align}
For $\alpha^{\perp,\upsilon'}_{m'n'} \neq 0$, we define $b = \omega_{c}-\alpha^{\perp,\upsilon'}_{m'n'}$ and change from $\omega$ to $x$  using $x=\omega-\alpha^{\perp,\upsilon'}_{m'n'}$:
\begin{align}
&I\left(\alpha^{\perp,\upsilon'}_{m'n'}\right) =-\dfrac{1}{\pi}\left(\dfrac{2g^2}{\kappa_{\perp}}\right)(\kappa_{\perp}/2)^2 \mathcal{P} \left[ \textstyle\int\limits_{-\alpha^{\perp,\upsilon'}_{m'n'}}^{\infty} dx \left(\dfrac{1}{x}\right)\left(\dfrac{1}{(\kappa_{\perp}/2)^2+(x-b)^2}\right) \right]
        \nonumber\\& \phantom{I\left(\alpha^{\perp,\upsilon'}_{m'n'}\right)} =-\dfrac{1}{\pi}\left(\dfrac{2g^2}{\kappa_{\perp}}\right)\dfrac{(\kappa_{\perp}/2)^2 }{(\kappa_{\perp}/2)^2+b^2}\mathcal{P}\left[\textstyle\int\limits_{-\alpha^{\perp,\upsilon'}_{m'n'}}^{\infty} dx\dfrac{1}{x} -\int\limits_{-\alpha^{\perp,\upsilon'}_{m'n'}}^{\infty} dx \dfrac{x-2b}{(\kappa_{\perp}/2)^2+(x-b)^2}\right]\nonumber
        \\& \phantom{I\left(\alpha^{\perp,\upsilon'}_{m'n'}\right)} =-\dfrac{1}{\pi}\left(\dfrac{2g^2}{\kappa_{\perp}}\right)\dfrac{(\kappa_{\perp}/2)^2 }{(\kappa_{\perp}/2)^2+b^2}\left[ \left.\ln\abs{x} \right\rvert_{-\alpha^{\perp,\upsilon'}_{m'n'}}^{\infty}  -\mathcal{P} \left(\int\limits_{-\alpha^{\perp,\upsilon'}_{m'n'}}^{\infty} dx \dfrac{x-2b}{(\kappa_{\perp}/2)^2+(x-b)^2}\right)\right]\,\vcenter{\hbox{.}}
\end{align}
For the first term in the square brackets, we switch back to $\omega$, 
\begin{align}
I\left(\alpha^{\perp,\upsilon'}_{m'n'}\right)  =-\dfrac{1}{\pi}\left(\dfrac{2g^2}{\kappa_{\perp}}\right)\dfrac{(\kappa_{\perp}/2)^2 }{(\kappa_{\perp}/2)^2+b^2}\left[ \left.\ln\abs{\omega-\alpha^{\perp,\upsilon'}_{m'n'}} \right\rvert_{0}^{\infty}  -\mathcal{P} \left(\int\limits_{-\alpha^{\perp,\upsilon'}_{m'n'}}^{\infty} dx \dfrac{x-2b}{(\kappa_{\perp}/2)^2+(x-b)^2}\right)\right]\,\vcenter{\hbox{.}}
    \label{eq_im1}
\end{align}
The principal value in the expression can be simplified by defining $u = x-b$:
\begin{align}
      \mathcal{P}\left[ \textstyle\int\limits_{-\alpha^{\perp,\upsilon'}_{m'n'}}^{\infty} dx \dfrac{x-2b}{(\kappa_{\perp}/2)^2+(x-b)^2} \right] &= \mathcal{P} \left[ \textstyle\int\limits_{-(\alpha^{\perp,\upsilon'}_{m'n'}+b)}^{\infty} du \dfrac{u}{(\kappa_{\perp}/2)^2+u^2}\right] -b \mathcal{P} \left[ \textstyle\int\limits_{-(\alpha^{\perp,\upsilon'}_{m'n'}+b)}^{\infty} du \dfrac{1}{(\kappa_{\perp}/2)^2+u^2} \right] \nonumber\\ &=\mathcal{P} \left[ \textstyle\int\limits_{-(\alpha^{\perp,\upsilon'}_{m'n'}+b)}^{\infty} du \dfrac{u}{(\kappa_{\perp}/2)^2+u^2}  -\left. \dfrac{b}{(\kappa_{\perp}/2)}\tan^{-1}\bm{\Biggl(}\dfrac{u}{(\kappa_{\perp}/2)}\bm{\Biggr)} \right\rvert_{-(\alpha^{\perp,\upsilon'}_{m'n'}+b)}^{\infty} \right]
      \nonumber\\&=\mathcal{P} \left[ \textstyle\int\limits_{-(\alpha^{\perp,\upsilon'}_{m'n'}+b)}^{\infty} du \dfrac{u}{(\kappa_{\perp}/2)^2+u^2} -\left. \dfrac{b}{(\kappa_{\perp}/2)}\tan^{-1}\bm{\Biggl(}\dfrac{u}{(\kappa_{\perp}/2)}\bm{\Biggr)} \right\rvert_{-(\alpha^{\perp,\upsilon'}_{m'n'}+b)}^{\infty} \right]\,\vcenter{\hbox{.}}
      \label{eq_Imint}
\end{align}
To evaluate the remaining integral in Eq.~(\ref{eq_Imint}), we define $t = u^2 + (\kappa_{\perp}/2)^2$ and use $dt = 2 u du$. This yields
\begin{align}
      {\mathcal{P}} \left[ \textstyle\int\limits_{-\alpha^{\perp,\upsilon'}_{m'n'}}^{\infty} dx \dfrac{x-2b}{(\kappa_{\perp}/2)^2+(x-b)^2} \right]&= \dfrac{1}{2}\mathcal{P} \left( \textstyle\int\limits_{\left(\alpha^{\perp,\upsilon'}_{m'n'}+b\right)^2 +\left(\sfrac{\kappa_{\perp}}{2}\right)^2}^{\infty} dt \dfrac{1}{t}\right) -\left.\dfrac{b}{(\kappa_{\perp}/2)}\tan^{-1}\bm{\Biggl(}\dfrac{u}{(\kappa_{\perp}/2)}\bm{\Biggr)} \right\rvert_{-(\alpha^{\perp,\upsilon'}_{m'n'}+b)}^{\infty} \nonumber\\&=\left.\dfrac{\ln\abs{t}}{2} \right\rvert_{\left(\alpha^{\perp,\upsilon'}_{m'n'}+b\right)^2 +\left(\sfrac{\kappa_{\perp}}{2}\right)^2}^{\infty} -\left. \dfrac{b}{(\kappa_{\perp}/2)}\tan^{-1}\bm{\Biggl(}\dfrac{u}{(\kappa_{\perp}/2)}\bm{\Biggr)} \right\rvert_{-(\alpha^{\perp,\upsilon'}_{m'n'}+b)}^{\infty} \,\vcenter{\hbox{.}}
      \nonumber
\end{align}
Reverting to the original variable $\omega$ and the corresponding limits, we obtain
\begin{align}
      {\mathcal{P}} \left[ \textstyle\int\limits_{-\alpha^{\perp,\upsilon'}_{m'n'}}^{\infty} dx \dfrac{\omega-2\omega_c+\alpha^{\perp,\upsilon'}_{m'n'}}{(\kappa_{\perp}/2)^2+(\omega-\omega_c)^2} \right] =  \left.\left[\dfrac{\ln\abs{(\omega-\omega_{c})^2+ (\kappa_{\perp}/2)^2}}{2}-\dfrac{\omega_{c}-\alpha^{\perp,\upsilon'}_{m'n'}}{(\kappa_{\perp}/2)}\tan^{-1}\bm{\Biggl(}\dfrac{\omega-\omega_{c}}{(\kappa_{\perp}/2)}\bm{\Biggr)}\right]\right\rvert_{0}^{\infty} \,\vcenter{\hbox{.}}
\label{eq_part1}
\end{align}
Substituting Eq.~(\ref{eq_part1}) into Eq.~(\ref{eq_im1}) and formally introducing the cutoff $\Upsilon$, we find 
\begin{align}        I\left(\alpha^{\perp,\upsilon'}_{m'n'}\right) =&-\dfrac{1}{\pi}\left(\dfrac{2g^2}{\kappa_{\perp}}\right)\dfrac{(\kappa_{\perp}/2)^2 }{(\kappa_{\perp}/2)^2+b^2}\lim_{\Upsilon\rightarrow \infty} \left.\left[\ln\abs{\omega-\alpha^{\perp,\upsilon'}_{m'n'}}-\dfrac{\ln\abs{(\omega-\omega_{c})^2+ (\kappa_{\perp}/2)^2}}{2}+\dfrac{\omega_{c}-\alpha^{\perp,\upsilon'}_{m'n'}}{(\kappa_{\perp}/2)}\tan^{-1}\bm{\Biggl(}\dfrac{\omega-\omega_{c}}{(\kappa_{\perp}/2)}\bm{\Biggr)}\right]\right\rvert_{0}^{\Upsilon} 
        \nonumber\\=&-\dfrac{1}{\pi}\left(\dfrac{2g^2}{\kappa_{\perp}}\right)\dfrac{(\kappa_{\perp}/2)^2 }{(\kappa_{\perp}/2)^2+b^2}\lim_{\Upsilon\rightarrow \infty} \left[\ln\abs{\Upsilon-\alpha^{\perp,\upsilon'}_{m'n'}}-\ln\abs{\alpha^{\perp,\upsilon'}_{m'n'}}-\dfrac{\ln\abs{(\Upsilon-\omega_{c})^2+ (\kappa_{\perp}/2)^2}}{2}+ \dfrac{\ln\abs{\omega^2_{b(c)}+ (\kappa_{\perp}/2)^2}}{2}\right]
        \nonumber\\&-\dfrac{1}{\pi}\left(\dfrac{2g^2}{\kappa_{\perp}}\right)\dfrac{(\kappa_{\perp}/2)^2 }{(\kappa_{\perp}/2)^2+b^2}\lim_{\Upsilon\rightarrow \infty} \left\{\dfrac{\omega_{c}-\alpha^{\perp,\upsilon'}_{m'n'}}{(\kappa_{\perp}/2)}\left[\tan^{-1}\bm{\Biggl(}\dfrac{\Upsilon-\omega_{c}}{(\kappa_{\perp}/2)}\bm{\Biggr)}-\tan^{-1}\bm{\Biggl(}\dfrac{-\omega_{c}}{(\kappa_{\perp}/2)}\bm{\Biggr)}\right]\right\}
        \nonumber\\=&\quad\,\dfrac{1}{\pi}\left(\dfrac{2g^2}{\kappa_{\perp}}\right)\dfrac{(\kappa_{\perp}/2)^2 }{(\kappa_{\perp}/2)^2+(\omega_{c}-\alpha^{\perp,\upsilon'}_{m'n'})^2}\left\{\ln\abs{\alpha^{\perp,\upsilon'}_{m'n'}}-\dfrac{\ln\abs{\omega^2_{c}+ (\kappa_{\perp}/2)^2}}{2}-\dfrac{\omega_{c}-\alpha^{\perp,\upsilon'}_{m'n'}}{(\kappa_{\perp}/2)}\left[\dfrac{\pi}{2}-\tan^{-1}\bm{\Biggl(}\dfrac{-\omega_{c}}{(\kappa_{\perp}/2)}\bm{\Biggr)}\right]\right\}
        \nonumber\\=&\;J_{\perp}(\alpha^{\perp,\upsilon'}_{m'n'})\left\{\ln\abs{\dfrac{\alpha^{\perp,\upsilon'}_{m'n'}}{\sqrt{\omega^2_{c}+ (\kappa_{\perp}/2)^2}}}-\dfrac{\omega_{c}-\alpha^{\perp,\upsilon'}_{m'n'}}{(\kappa_{\perp}/2)}\left[\dfrac{\pi}{2}-\tan^{-1}\bm{\Biggl(}\dfrac{-\omega_{c}}{(\kappa_{\perp}/2)}\bm{\Biggr)}\right]\right\}\,\vcenter{\hbox{.}}
\end{align}
For $\alpha^{\perp,\upsilon'}_{m'n'}=0$, the term $\ln\abs{\alpha^{\perp,\upsilon'}_{m'n'}}$ diverges logarithmically. This divergence can be dealt with by regularization, followed by subtracting the divergent part. This procedure yields
\begin{align}
   I\left(0\right) =J_{\perp}(0)\left\{\ln\abs{\dfrac{(\kappa_{\perp}/2)}{\sqrt{\omega^2_{c}+ (\kappa_{\perp}/2)^2}}}-\dfrac{\omega_{c}}{(\kappa_{\perp}/2)}\left[-\dfrac{\pi}{2}-\tan^{-1}\bm{\Biggl(}\dfrac{-\omega_{c}}{(\kappa_{\perp}/2)}\bm{\Biggr)}\right]\right\}\,\vcenter{\hbox{.}}
\end{align}
To extract the decay rates and Lamb shifts from the dissipation coefficients in Eq.~(\ref{eq_disscoffcav}), 
we write 
\begin{equation}
    \exp[i\left(\omega^{\perp,\upsilon}_{d}-\omega^{\perp,\upsilon'}_{d}\right)t] = \cos\left(\delta^{\perp,\upsilon\upsilon'}_{d}t\right)+i\sin\left(\delta^{\perp,\upsilon\upsilon'}_{d}t\right)\,.
\end{equation}
This allows us to write
\begin{equation}
    \Gamma_{\perp,mn,m'n'}(t) = \smashoperator{\sum_{\substack{\upsilon\,\in\{12,21\} \\ \upsilon'\in\{12,21\}}}}  \xi^{\perp,\upsilon}_{mn}\xi^{\perp,\upsilon'}_{m'n'} \left\{\left[\cos\left(\delta^{\perp,\upsilon\upsilon'}_{d}t\right)R\left(\alpha^{\perp,\upsilon'}_{m'n'}\right)-\sin\left(\delta^{\perp,\upsilon\upsilon'}_{d}t\right)I\left(\alpha^{\perp,\upsilon'}_{m'n'}\right)\right] + i\left[\cos\left(\delta^{\perp,\upsilon\upsilon'}_{d}t\right)I\left(\alpha^{\perp,\upsilon'}_{m'n'}\right)+\sin\left(\delta^{\perp,\upsilon\upsilon'}_{d}t\right)R\left(\alpha^{\perp,\upsilon'}_{m'n'}\right)\right]\right\}\,.
\end{equation}
The real part of the dissipation coefficients provides the decay rates $\gamma_{\perp,mn,m'n'}(t)$,
\begin{equation}
    \gamma_{\perp,mn,m'n'}(t) =  \smashoperator{\sum_{\substack{\upsilon\,\in\{12,21\} \\ \upsilon'\in\{12,21\}}}}  \xi^{\perp,\upsilon}_{mn}\xi^{\perp,\upsilon'}_{m'n'} \left[\cos\left(\delta^{\perp,\upsilon\upsilon'}_{d}t\right)R\left(\alpha^{\perp,\upsilon'}_{m'n'}\right)-\sin\left(\delta^{\perp,\upsilon\upsilon'}_{d}t\right)I\left(\alpha^{\perp,\upsilon'}_{m'n'}\right)\right]
    \label{eq_decayrate_perp}
\end{equation}
while the imaginary part of the dissipation coefficients provides the Lamb shifts $\Delta_{\perp,mn,m'n'}(t)$, 
\begin{equation}
    \Delta_{\perp,mn,m'n'}(t) =  \smashoperator{\sum_{\substack{\upsilon\,\in\{12,21\} \\ \upsilon'\in\{12,21\}}}} \xi^{\perp,\upsilon}_{mn}\xi^{\perp,\upsilon'}_{m'n'} \left[\cos\left(\delta^{\perp,\upsilon\upsilon'}_{d}t\right)I\left(\alpha^{\perp,\upsilon'}_{m'n'}\right)+\sin\left(\delta^{\perp,\upsilon\upsilon'}_{d}t\right)R\left(\alpha^{\perp,\upsilon'}_{m'n'}\right)\right]\,.
\end{equation}
\end{widetext}
The decay rates $\gamma_{\perp,mn,m'n'}(t)$ can further be simplified by applying the RWA, which is valid provided $\omega_p \gg \Gamma_{\perp}$.
Within the RWA, the time dependence drops out and we have
\begin{equation}
    \gamma_{\perp,mn,m'n'} =   \xi^{\perp,12}_{mn}\xi^{\perp,12}_{m'n'} R\left(\alpha^{\perp,12}_{m'n'}\right)\,.
\end{equation}

The Lamb shifts, which are not accounted for in the numerical results presented in our paper, enter into the coherent master equation dynamics. Since the Lamb shifts are small, they may change the dynamics quantitatively but not qualitatively. The decay rates, in contrast, contribute to the incoherent time evolution of the master equation. Their inclusion leads to qualitative changes of the dynamics in certain parameter regimes. 
 
Similarly, we calculate the dissipation coefficients for the bath. Using Eqs.~(\ref{eq_lambath}) and (\ref{eq_specbath}), we find
\begin{align}
    \Gamma_{\parallel,mn,m'n'} &= \xi^{\parallel}_{mn}\xi^{\parallel}_{m'n'}\left[R\left(\alpha^{\parallel}_{m'n'}\right) + iI\left(\alpha^{\parallel}_{m'n'}\right)\right]\,,
    \nonumber
\end{align}
\begin{align}
    R\left(\alpha^{\parallel}_{m'n'}\right) &=
    \begin{cases}
    \pi J_{\parallel}\left(\alpha^{\parallel}_{m'n'}\right)\,, &\alpha^{\parallel}_{m'n'}\geq 0  
    \\0\,, &\alpha^{\parallel}_{m'n'}<0  
    \end{cases}
    \nonumber\\ 
    I\left(\alpha^{\parallel}_{m'n'}\right) &=
    \begin{cases}
    J_{\parallel}(\alpha^{\parallel}_{m'n'})\left[\ln\abs{\dfrac{\alpha^{\parallel}_{m'n'}}{({\kappa_{\parallel}}/2)}}+\dfrac{\alpha^{\parallel}_{m'n'}}{({\kappa_{\parallel}}/2)}\dfrac{\pi}{2}\right]\,\vcenter{\hbox{,}} &\alpha^{\parallel}_{m'n'}\neq 0  
    \\0\,, &\alpha^{\parallel}_{m'n'}=0 
    \end{cases}\,\raisebox{-0.4ex}{.}
\end{align}
After some work, we find 
\begin{align}
    \begin{split}
        &\gamma_{\parallel,mn,m'n'} =  \xi^{\parallel}_{mn}\xi^{\parallel}_{m'n'} R\left(\alpha^{\parallel}_{m'n'}\right)\,,
        \\&\Delta_{\parallel,mn,m'n'} = \xi^{\parallel}_{mn}\xi^{\parallel}_{m'n'} I\left(\alpha^{\parallel}_{m'n'}\right)
    \end{split}
    \label{eq_decayrate_par}
\end{align}
for the dephasing rates $\gamma_{\parallel,mn,m'n'}$ and the Lamb shifts $\Delta_{\parallel,mn,m'n'}$, respectively. Note that both the dephasing rates and the Lamb shifts are, just as the corresponding dissipation coefficients, independent of time. 

\section{\label{sec:appendD}Quantum regression theorem}
This appendix 
is devoted to the evaluation of two-time correlation functions used in the resonance-fluorescence spectra via the quantum regression theorem, following standard treatments~\cite{steck_quantum_atom_optics}. Rather than working in the interaction picture, we work in the Schrödinger and Heisenberg pictures throughout. Transformations between these pictures are implemented using the unitary time-evolution operator $U_T(t,t')$, which propagates the full qutrit--cavity--bath composite system from time $t'$ to time $t$. In this appendix, $\rho$ refers to the density matrix in either the laboratory frame or the rotating frame (the derivation is the same for both frames).

The emission/absorption spectrum $S_{ab}(\omega)$ is defined as the Fourier-transform of the two-time correlator 
$g_{ab}(t,\tau)$,
\begin{equation}
    S_{ab}(\omega)  = \lim_{t \rightarrow \infty} \left[\dfrac{1}{2\pi}\int_{-\infty}^{\infty} d\tau e^{-i\omega \tau} g_{ab}(t,\tau)\right]\,,
\end{equation}
where the correlation function (in the Heisenberg picture) reads
\begin{equation}
g_{ab}(t,\tau)=\expval{\delta\sigma^{\dagger}_{ab}(t)\delta\sigma_{ab}(t + \tau)} = \mathrm{Tr}\left[\delta\sigma^{\dagger}_{ab}(t)\delta\sigma_{ab}(t + \tau)\rho\right]\,,
\end{equation}
where $\delta\sigma_{mn} (t) = \sigma_{mn} (t) -\expval{\sigma_{mn}}_{ss}$, $\sigma_{mn} = \ket{m}\bra{n}$, and $\expval{\sigma_{mn}}_{ss}$ is the steady state expectation value of the operator. Tr stands for the trace over the qutrit, the cavity, and the bath variables and $\rho$ is the composite density operator, which is, in the Heisenberg picture, fixed at the initial value. The reduced density matrices, obtained after partial traces over the qutrit, the cavity, and the bath, are 
\begin{equation}
          \rho_q = \mathrm{Tr}_{c,b}\left(\rho\right)\,, 
    \quad \rho_c = \mathrm{Tr}_{q,b}\left(\rho\right)\,, 
    \quad \rho_b = \mathrm{Tr}_{q,c}\left(\rho\right)\,.
\end{equation}
For $E_a<E_{b}$ and $E_a>E_{b}$, 
$S_{ab}(\omega)$ corresponds to an emission spectrum and an absorption spectrum, respectively. 

The two-time correlator $g_{ab}(t,\tau)$ is computed using the quantum regression theorem~\cite{shavit2019,lax1963}. To this end, the operators in the correlation function are expressed in the Schr\"odinger picture by means of the unitary time-evolution operator $U_T(t,0)$, which propagates the composite qutrit-cavity-bath system from time $0$ to time $t$,
\begin{align}
\begin{split}
\label{eq_appendixD_1}
    \delta\sigma^{\dagger}_{ab}(t) &= U^{\dagger}_T(t,0)\delta\sigma^{\dagger}_{ab}U_T(t,0)\,, \\
    \delta\sigma_{ab}(t+\tau) &= U^{\dagger}_T(t+\tau,0)\delta\sigma_{ab}U_T(t+\tau,0)\,.
\end{split}
\end{align}
Note that $U_T(t,t')$ is distinct from the operators $U(t)$, $U_{q}(t)$, and $U_{b}(t)$. In the Schr\"odinger picture, the correlation function reads
\begin{flalign}
&g_{ab}(t,\tau)
\nonumber\\&=\mathrm{Tr}\left[U^{\dagger}_T(t,0)\delta\sigma^{\dagger}_{ab}U_T(t,0)U^{\dagger}_T(t+\tau,0)\delta\sigma_{ab}U_T(t+\tau,0)\rho(0)\right]
\nonumber\\&= \mathrm{Tr}\left[\delta\sigma^{\dagger}_{ab}U_T(t,0)U^{\dagger}_T(t+\tau,0)\delta\sigma_{ab}U_T(t+\tau,0)\rho(0) U^{\dagger}_T(t,0)\right] \,.&&
\end{flalign}
To obtain the expression after the last equal sign, we used the cyclic property of the trace. The correlation function can be further simplified using the following properties of the evolution operator:
\begin{equation}
\label{eq_timeoperator}
    U_T(t,t')U_T(t',t'') = U_T(t,t''),\quad U^{\dagger}_T(t,t') = U_T(t',t).
\end{equation}
Using Eq.~(\ref{eq_timeoperator}), we find
\begin{align}
    &g_{ab}(t,\tau)  
    \nonumber\\ &= \mathrm{Tr}\left[\delta\sigma^{\dagger}_{ab}U^{\dagger}_T(t+\tau,t)\delta\sigma_{ab}U_T(t+\tau,0)\rho(0) U^{\dagger}_T(t,0)\right]
    \nonumber\\ &= \mathrm{Tr}\left[\delta\sigma^{\dagger}_{ab}U^{\dagger}_T(t+\tau,t)\delta\sigma_{ab}U_T(t+\tau,0)U^{\dagger}(t,0)U(t,0)\rho(0) U^{\dagger}_T(t,0)\right]
    \nonumber\\ &= \mathrm{Tr}\left[\delta\sigma^{\dagger}_{ab}U^{\dagger}_T(t+\tau,t)\delta\sigma_{ab}U_T(t+\tau,t)\{U(t,0)\rho(0) U^{\dagger}_T(t,0)\}\right]\,.
\end{align}
Introducing the time-dependent composite density matrix $\rho(t)$ in the Schr\"odinger picture, $\rho(t) = U_T(t,0)\rho(0) U^{\dagger}_T(t,0)$, and again using the cyclic property of the trace, we find
\begin{flalign}
    g_{ab}(t,\tau) &=\mathrm{Tr}\left[\delta\sigma^{\dagger}_{ab}U^{\dagger}_T(t+\tau,t)\delta\sigma_{ab}U_T(t+\tau,t)\rho(t) \right]
    \nonumber\\
    &=\mathrm{Tr}\left[\delta\sigma_{ab}U_T(t+\tau,t)\rho(t)\delta\sigma^{\dagger}_{ab} U^{\dagger}_T(t+\tau,t)\right]
    \nonumber\\
    &=\mathrm{Tr}_{q}\left[\delta\sigma_{ab}\mathrm{Tr}_{c,b}\left[U_T(t+\tau,t)\rho(t)\delta\sigma^{\dagger}_{ab} U^{\dagger}_T(t+\tau,t)\right]\right]\,.&&
\end{flalign}
In the last equality, we used that the trace operation can be split, namely, we used $\text{Tr}=\text{Tr}_q \text{Tr}_{c,b}$, and that $\delta\sigma_{ab}$ only acts on the qutrit. 
To proceed, we define the two-time operator $\Lambda(t+\tau,t)$, 
\begin{equation}
    \Lambda(t+\tau,t) = \mathrm{Tr}_{c,b}\left[U_T(t+\tau,t)\rho(t)\delta\sigma^{\dagger}_{ab}U^{\dagger}_T(t+\tau,t)\right]\,.
\end{equation}
The operator $\Lambda(t+\tau,t)$ obeys the same equation of motion (master equation) as the reduced density matrix $\rho_q(t)$ of the qutrit, with the same Liouvillian generator ${\mathcal{L}}$. However, while the derivative is taken with respect to time $t$ for the density matrix evolution, it is taken with respect to the delay time $\tau$ for the two-time operator evolution: 
\begin{equation}
    \frac{d}{d t} \rho_q(t) = \mathcal{L}\rho_q(t) \implies \frac{d}{d \tau} \Lambda(t+\tau,t) = \mathcal{L}\Lambda(t+\tau,t)\,.
\end{equation}
The initial condition is given by
\begin{align}
    \Lambda(t,t)  &= \mathrm{Tr}_{c,b}\left[U_T(t,t)\rho(t)\delta\sigma^{\dagger}_{ab}U^{\dagger}_T(t,t)\right] 
    \nonumber\\&= \mathrm{Tr}_{c,b}\left[\rho(t)\right]\delta\sigma^{\dagger}_{ab}
    \nonumber\\&= \rho_q(t)\delta\sigma^{\dagger}_{ab}\,.
\end{align}
With this, the correlation function reduces to 
\begin{equation}
    g_{ab}(t,\tau)  =\mathrm{Tr}_q\left[\delta\sigma_{ab}\Lambda(t+\tau,t)\right]\,.
\end{equation}
 For emission or absorption spectra, correlation functions are typically computed in the long time limit, i.e.,
\begin{equation}
    \lim_{t\rightarrow \infty}g_{ab}(t,\tau)  =\mathrm{Tr}_q\left[\delta\sigma_{ab}\Lambda(\tau)\right]\,,
\end{equation}
where 
\begin{equation}
    \frac{d}{d \tau} \Lambda(\tau) = \mathcal{L}\Lambda(\tau), \quad \Lambda(0) = \rho_q(t\rightarrow \infty)\delta\sigma^{\dagger}_{ab}\,.
\end{equation} 
In practice, we compute the spectrum in the rotating frame. 

\section{\label{sec:appendE}Interpretation of the spectrum}
This appendix is dedicated to understanding the fluroscence spectrum $S_{12}(\omega)$ as well as to elucidating the physical origin of the deviations observed when computing $S_{12}(\omega)$ via different master-equation frameworks. Fluorescence from a coherently driven quantum system consists of a coherent (elastic) component and an incoherent (inelastic) component. The coherent contribution, which does not carry any information about dissipative processes, results in a ``trivial'' delta-function peak at the drive frequency. The incoherent part, in contrast, contains ``non-trivial'' features such as the Mollow triplets~\cite{mollow1969}, which reflect the interplay between coherent driving and system--bath coupling. This appendix provides an explicit discussion of the fluorescence spectrum for Application 1; Application~2 can be analyzed analogously.

Since our aim in this appendix is to explain the key features of the fluorescence spectra of the three-level system over the entire parameter regime, the analysis that follows is based on the Liouvillian generator ${\cal{L}}$.
Specifically, we use the Liouvillian generator ${\cal{L}}$ \cite{ivar2004,Yanay2020} to rewrite Eq.~(\ref{eq_UME}) compactly, 
\begin{eqnarray}
\label{eq_superoperator}
    \frac{d}{dt} \rho_{q,R}(t) = {\cal{L}} \rho_{q,R}(t)\,.
\end{eqnarray}
Note that the Liouvillian generator ${\cal{L}}$ depends on whether we are considering the IME, the rotating-frame master equation, or the laboratory-frame master equation. 

In the absence of the $(2{\leftrightarrow}3)$-drive, state $\ket{3}$ is only coupled dissipatively. As a consequence, states $\ket{1}$ and $\ket{2}$ are linear combinations of the dressed states $|\mu_+\rangle$ and $|\mu_-\rangle$ (i.e., $|\mu_0\rangle$ does not contribute). It follows that the 21-fluorescence spectrum can be fully explained by considering the states $|\mu_+\rangle$ and $|\mu_-\rangle$ since other density matrix elements have zero overlap with the emission operator.
For the analysis of the 21-fluorescence spectrum, it is thus sufficient to restrict the sums in Eq.~(\ref{eq_UME}) over $m,n,m',n'$ to $+$ and $-$. Correspondingly, the density matrix in Eq.~(\ref{eq_superoperator}) reduces, for the purpose of interpreting the 21-fluorescence spectrum, to a $2 \times 2$ matrix with elements 
$\rho_{q,R,++}$, $\rho_{q,R,+-}$, $\rho_{q,R,-+}$, and $\rho_{q,R,--}$.
Arranging these four matrix elements as a four-component vector $\vec{\rho}_{q,R}$ and, correspondingly, arranging the relevant elements of the generator into the $4 \times 4$ matrix ${M}$, we calculate the eigenvalues $\lambda_j$ and eigenvectors $\vec{V}_j$ of ${M}$. 
The eigenvalues are
\begin{align}
      &\lambda_0=0\,,
    \\&\lambda_1 = -\gamma_0\,,
    \\&\lambda_2 = -\gamma_s -i\Omega_{\Gamma}\,,
    \\&\lambda_3 = -\gamma_s +i\Omega_{\Gamma}\,,
\end{align}
where $\gamma_0$, $\gamma_s$, and $\Omega_{\Gamma} $ are real. 
The eigenvalue $\lambda_0$ determines the steady-state density matrix. Specifically, $\vec{\rho}_{q,R}^{\,ss}$ is equal to $\vec{V}_0$.

The eigenvectors $\vec{V}_1$, $\vec{V}_2$, and $\vec{V}_3$ are referred to as dynamical modes~\cite{Yanay2020}. These dynamical modes and their eigenvalues govern the fluorescence spectrum. The fact that there exist three dynamical modes implies that the spectrum consists of three peaks. We find that the spectrum can be written as 
\begin{align}
\label{eq_spectrum_analytical}
     S_{12}(\omega) = \smashoperator{\sum_{j =1,2,3}} \dfrac{\Re(C_j)\Re(\lambda_j)+\Im(C_j)\left[\omega+\Im(\lambda_j)\right]}{\left[\Re(\lambda_j)\right]^2 + \left[\omega+\Im(\lambda_j)\right]^2}\,\vcenter{\hbox{,}} 
\end{align}
where the real and imaginary parts of $C_j$ depend on the elements of the dynamical mode $\vec{V}_j$, the steady-state density matrix elements, and the matrices $\xi^{\perp,12}$ and $\xi^{\perp,21}$ [see Eq.~(\ref{eq_xiperp})], 
which encode the overlap of the dressed states with the transition operator that is associated with the fluorescence spectrum.
The real part of the ``weight factor'' $C_j$ determines the amplitude of the purely Lorentzian peak, whereas the imaginary part produces an antisymmetric dispersive component that modifies the otherwise symmetric spectral profile. Explicit analytical expressions for the coefficients $C_j$ will be provided in a forthcoming publication~\cite{BasakInPrep}. Equation~(\ref{eq_spectrum_analytical}) shows that the real part of the eigenvalues $\lambda_j$ ($j=1-3$) sets the decoherence rate and thus the width of the peak that is associated with the $j$th dynamical mode. The imaginary part of the eigenvalues sets the oscillation frequency and thus the position of the peak. Since $\lambda_1$ is purely real, the peak that originates from the dynamical mode $\vec{V}_1$ is centered at $\omega=0$.

\begin{figure}
    \centering
    \includegraphics[width=3.375in]{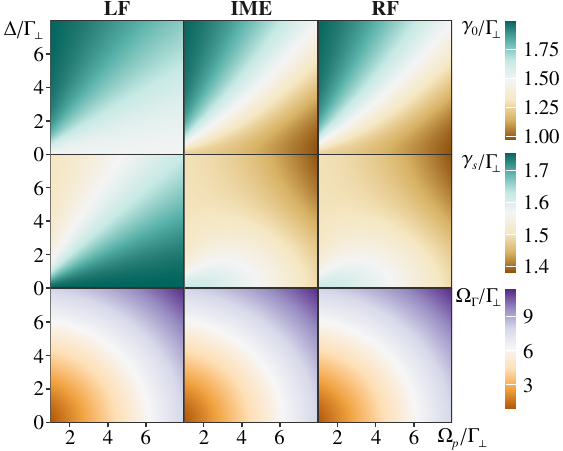}
    \caption{\label{fig:exponents}(color online) The real and imaginary parts of the eigenvalues $\lambda_1,\lambda_2$, and $\lambda_3$, which determine the widths and positions, respectively, of the spectral peaks, are shown as functions of the scaled coupling $\Omega_p/\Gamma_{\perp}$ and the scaled detuning $\Delta/\Gamma_{\perp}$. The first row shows $\gamma_0$, the second row shows $\gamma_s$, and the third row shows $\Omega_{\Gamma}$. The quantities are calculated for the laboratory-frame (LF) master equation (column 1), the IME (column 2), and the rotating-frame (RF) master equation (column 3). The other parameters are the same as those used in Fig.~\ref{fig:App1RS12} for the top set of spectra, namely the spectra with $\Gamma_{\parallel}/\Gamma_{\perp} = 0.5$: $\Omega_s= 0$, $\kappa_\perp = \kappa_{\parallel} = 30\Gamma_{\perp}$, $\omega_{\perp} = \omega_{1}=3\times 10^5 \Gamma_{\perp},\,\omega_{\parallel} =0,\,\omega_1-\omega_3 = 10\Gamma_{\perp},\,\text{and } T_b =0$.}
\end{figure}

\begin{figure}
    \centering
    \includegraphics[width=3.375in]{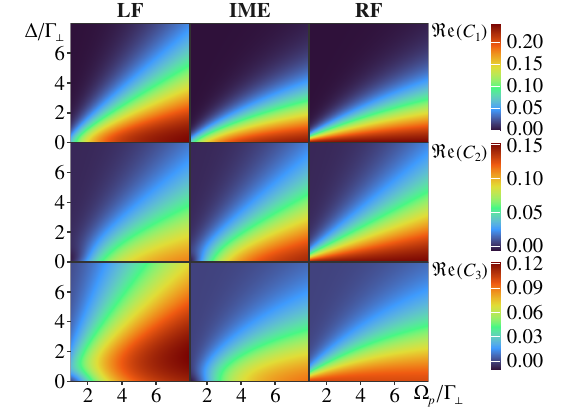}
    \caption{\label{fig:symmheight}(color online) The real part of the weight factors $C_1$ (row 1), $C_2$ (row 2), and $C_3$ (row 3), which determine the weights of the purely Lorentzian contribution to the peaks centered at $\omega = 0$, $\omega = \Omega_{\Gamma}$, and $\omega = -\Omega_{\Gamma}$, respectively. The results are shown as functions of the scaled coupling $\Omega_p/\Gamma_{\perp}$ and the scaled detuning $\Delta/\Gamma_{\perp}$. The quantities are calculated for the laboratory-frame (LF) master equation (column 1), the IME (column 2), and the rotating-frame (RF) master equation (column 3). The other parameters are the same as those used in Fig.~\ref{fig:App1RS12} for the top set of spectra, namely the spectra with $\Gamma_{\parallel}/\Gamma_{\perp} = 0.5$: $\Omega_s= 0$, $\kappa_\perp = \kappa_{\parallel} = 30\Gamma_{\perp}$, $\omega_{\perp} = \omega_{1}=3\times 10^5 \Gamma_{\perp},\,\omega_{\parallel} =0,\,\omega_1-\omega_3 = 10\Gamma_{\perp},\,\text{and } T_b =0$.} 
\end{figure}

\begin{figure}
    \centering
    \includegraphics[width=3.375in]{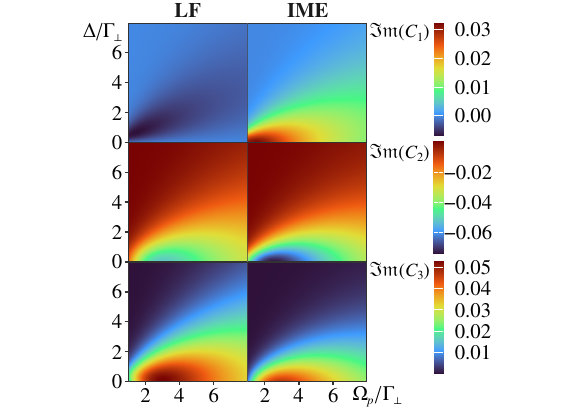}
    \caption{\label{fig:dispersive}(color online) The imaginary part of the weight factors $C_1$ (row 1), $C_2$ (row 2), and $C_3$ (row 3), which determine the antisymmetric contribution to the peaks centered at $\omega = 0$, $\omega = \Omega_{\Gamma}$, and $\omega = -\Omega_{\Gamma}$ respectively. The results are shown as a function of the scaled coupling $\Omega_p/\Gamma_{\perp}$ and the scaled detuning $\Delta/\Gamma_{\perp}$. The quantities are calculated for the laboratory-frame (LF) master equation (column 1) and the IME (column 2). For the rotating-frame master equation, the imaginary parts of $C_1$, $C_2$, and $C_3$ are identically zero. The other parameters are the same as those used in Fig.~\ref{fig:App1RS12} for the top set of spectra, namely the spectra with $\Gamma_{\parallel}/\Gamma_{\perp} = 0.5$: $\Omega_s= 0$, $\kappa_\perp = \kappa_{\parallel} = 30\Gamma_{\perp}$, $\omega_{\perp} = \omega_{1}=3\times 10^5 \Gamma_{\perp},\,\omega_{\parallel} =0,\,\omega_1-\omega_3 = 10\Gamma_{\perp},\,\text{and } T_b =0$.}
\end{figure}

To show the dependence of the quantities $\gamma_0$, $\gamma_s$, $\Omega_{\Gamma}$, $\Re{(C_1)}$, $\Re{(C_2)}$, $\Re{(C_3)}$, $\Im{(C_1)}$, $\Im{(C_2)}$, and $\Im{(C_3)}$, which govern the fluorescence spectrum $S_{12}(\omega)$, on the master equation framework employed, we vary $\Delta/\Gamma_{\perp}$ and $\Omega_p/\Gamma_{\perp}$ and set the other parameters to be the same as those employed in Fig.~\ref{fig:App1RS12}; specifically, we focus on the top set of spectra in Figs.~\ref{fig:App1RS12} and \ref{fig:App1RS12_U}, i.e., we set $\Gamma_{\parallel}/\Gamma_{\perp}=0.5$. Figures~\ref{fig:exponents}, \ref{fig:symmheight}, and \ref{fig:dispersive} show the results. 

The third column of Fig.~\ref{fig:exponents} shows that $\Omega_T/\Gamma_{\perp}$ is, for the parameter combinations considered, essentially the same for all three master-equation frameworks considered (columns 1, 2, and 3 are for the laboratory-frame master equation, the IME, and the rotating-frame master equation, respectively). Since $\Omega_{\Gamma}$ governs the positions of the side peaks, this explains why the side peak positions in Figs.~\ref{fig:App1RS12} and \ref{fig:App1RS12_U} are essentially independent of the master-equation framework employed. The first and second row of Fig.~\ref{fig:exponents} show that the decoherence rates $\gamma_0$ and $\gamma_s$, which govern the widths of the central and side peaks, are approximately the same for the IME (column 2) and the rotating-frame master equation (column 3), but differ for the laboratory-frame master equation (column 1). Correspondingly, the peak widths of the spectra obtained within the laboratory-frame master equation differ slightly from those obtained within the IME and the rotating-frame master equation. These changes are not visible on the scale of Figs.~\ref{fig:App1RS12} and \ref{fig:App1RS12_U} since the differences in $\gamma_0$ and $\gamma_s$ for the laboratory-frame master equation and for the other two master equations are small compared to the range of $\omega$ values considered in Figs.~\ref{fig:App1RS12} and \ref{fig:App1RS12_U} (namely, $\omega/\Gamma_{\perp}\in[-12,12]$).

Figure~\ref{fig:symmheight} shows the real part of the weight factors $C_j$, which determine the amplitude of the Lorentzian contributions to the spectral peaks. For comparatively small values of the coupling $\Omega_p/\Gamma_{\perp}$ and the detuning $\Delta/\Gamma_{\perp}$ (lower left corner of the plots), the results obtained within the laboratory-frame master equation and the IME agree quite well, while the results obtained within the rotating-frame master equation deviate. If either $\Omega_p/\Gamma_{\perp}$ or $\Delta/\Gamma_{\perp}$ are ``large'' or both (i.e., away from the lower left corner of the plots), the results obtained within the rotating-frame master equation and the IME agree quite well, while the results obtained within the laboratory-frame master equation deviate. This reflects the fact that the laboratory-frame master equation correctly captures the behavior in the weak-driving regime, while the rotating-frame master equation is more accurate for stronger driving or larger detuning. 
The limitations of the laboratory-frame master equation arise from assigning the same bath response to the different dressed-state transitions, even though these transitions occur at different energies. This becomes particularly significant for dephasing processes that describe energy exchange with the bath at frequencies $\omega = 0$ and $\omega = \pm \Omega_T$. 
Because the laboratory-frame master equation approach assigns the same bath response to these processes, it fails to capture the suppression of transitions at negative frequencies (i.e., at $\omega \approx -\Omega_T$); such a suppression should exist since the zero-temperature bath cannot supply energy.

The limitations of the rotating-frame master equation become evident when examining the imaginary part of the weight factors $C_j$ (see Fig.~\ref{fig:dispersive}). Specifically, the imaginary part of the $C_j$, which determine the dispersive contribution to the spectrum, vanish. This means that the rotating-frame master equation yields a purely Lorentzian spectrum with vanishing dispersive contributions. 
This stems from the structure of the matrix $M$ within the rotating-frame master equation, where the ``population sector'' and ``coherence sector'' are not coupled to each other (i.e., where $\rho_{q,R,++}$ and $\rho_{q,R,--}$ are decoupled from $\rho_{q,R,+-}$ and $\rho_{q,R,-+}$). 
Figure~\ref{fig:dispersive} shows that the imaginary parts of the $C_j$ are non-zero for the laboratory-frame master equation and the IME. While the overall dependence of $\Im{(C_j)}$ on $\Delta/\Gamma_{\perp}$ and $\Omega_p/\Gamma_{\perp}$ is similar for these two approaches, differences are visible.

The analysis of the spectra for Application 1 presented in the section can be extended to Application 2~\cite{BasakInPrep}. A key take-away message is that the developments presented in our 
paper not only allow us to treat driven $N$-level systems in previously inaccessible parameter regimes but also provide a powerful framework for interpreting observables, such as the fluorescence spectrum, in a transparent manner.
\vspace{-1em}
\bibliography{references}
\end{document}